\documentclass[aps,prd,floats,nofootinbib,preprintnumbers]{revtex4}
\usepackage{epsfig}
\usepackage{amsmath}
\usepackage{amssymb}
\usepackage{amsthm}
\usepackage{longtable}

\begin{document}

\preprint{NUHEP-TH/08-02}

\title{The Physical Range of Majorana Neutrino Mixing Parameters}

\author{Andr\'e de Gouv\^ea}
\affiliation{Northwestern University, Department of Physics \&
Astronomy, 2145 Sheridan Road, Evanston, IL~60208, USA}

\author{James Jenkins}
\affiliation{Northwestern University, Department of Physics \&
Astronomy, 2145 Sheridan Road, Evanston, IL~60208, USA}

\begin{abstract}
If neutrinos are Majorana fermions, the lepton mixing parameter space consists of six mixing parameters: three mixing angles and three CP-odd phases. A related issue concerns the physical range of the mixing parameters. What values should these take so that all physically distinguishable mixing scenarios are realized? We present a detailed discussion of the lepton mixing parameter space in the case of two and three active neutrinos, and in the case of three active and $N$ sterile neutrinos. We emphasize that this question, which has been a source of confusion even among ``neutrino'' physicists, is connected to an unambiguous definition of the neutrino mass eigenstates. We find that all Majorana phases can always be constrained to lie between $0$ and $\pi$, and that all mixing angles can be chosen positive and at most less than or equal to $\pi/2$ provided the Dirac phases are allowed to vary between $-\pi$ and $\pi$. We illustrate our results with several examples. Finally, we point out that, in the case of new flavor-changing neutrino interactions, the lepton mixing parameter space may need to be enlarged. We properly qualify this statement, and offer concrete examples. 
\end{abstract}

\maketitle

\setcounter{equation}{0} \setcounter{footnote}{0}
\section{Introduction}
\label{sec:Intro}

It is now established that, similar to quarks, leptons mix \cite{NeutrinoReview}. The phenomenon of fermion mixing was identified long ago (see, for example, \cite{Cabibbo:1963yz,Branco:1999fs}). It is a consequence of the fact that we have (at least) three sets of each ``type'' of chiral fermion field ($Q_L,L_L,e_R,u_R,$\ldots) and, while gauge interactions cannot tell different generations apart, the interactions responsible for fermion masses can. 

In the quark sector of the standard model, it is well-known that the ambiguity in defining the different quark chiral fields allows one to describe the entire flavor sector of the quarks in terms of 10 parameters: six quark masses (three for the up-type quarks, three for the down-type quarks) and four real parameters in the Cabibbo-Kobayashi-Maskawa matrix (say, the Wolfenstein parameters $\lambda=\sin\theta_C$, where $\theta_C$ is the Cabibbo angle, $A,\rho,\eta$). It is also well-known (but often not emphasized) that one can choose all quark masses positive and all quark mixing angles to lie in the first quadrant (for the Wolfenstein parameters, this translates into $\lambda>0$, $A>0$).

In the lepton sector the situation is similar, with a couple of ``twists:'' the neutrinos may be Majorana fermions, and the neutrino sector of the Standard Model can be easily enlarged to included so-called sterile neutrinos $\nu_{s}$. Unlike fourth-generation quarks or charged leptons, sterile neutrinos can arise from Standard-Model singlet chiral fermion fields, and mix with the so-called active neutrinos after electroweak symmetry is broken. Not only are sterile neutrinos of virtually all masses allowed by all current neutrino data, they are natural consequences of several well-motivated scenarios that explain the origin of the Majorana neutrino masses, including most versions of the  seesaw mechanism \cite{SeeSaw}. It has recently been emphasized that there are plausible reasons to expect the ``seesaw'' sterile neutrinos to be low-mass, propagating fields \cite{low_seesaw}.

In the absence of sterile neutrinos, it is well-known \cite{maj_CP_phase,Schechter:1980gr} that the flavor sector of the leptons is described in terms of 12 parameters: three charged-lepton masses, three neutrino masses, and six real parameters in the lepton mixing matrix, often referred to as the Pontecorvo-Maki-Nakagawa-Sakata matrix. The last are parameterized in terms of three real mixing angles ($\theta_{12},\theta_{23},\theta_{13}$) and three complex phases. If the neutrinos are Dirac fermions, two of the three complex phases are unphysical, while if one of the neutrino masses is identically zero, one of the three complex phases can be ``rotated away.'' 

In this paper, we discuss the physical range of the lepton mixing parameters (mixing angles and complex phases) assuming that the neutrinos are Majorana fermions and allowing for the presence of sterile neutrinos. By ``physical range'' we refer to the following question: `given a proper definition of the different neutrino states, what values should the mixing parameters take in order to describe {\sl all} physically distinguishable values of {\sl all} observables?'. Subsets of this discussion, usually concentrating on the mixing angles and the ``Dirac'' phase, can be found elsewhere (see, for example, \cite{Branco:1999fs,Fogli:1996ne,de Gouvea:2000cq,GonzalezGarcia:2000sq,DeGouvea:2001ag,Fornengo:2001pm,Latimer:2004hd,Yao:2006px,Jenkins:2007ip} for an incomplete list of references), and many of the results we present here in detail can be found throughout the literature. It is, however, our experience that the topic of the ``physical" lepton-mixing parameter space is often source of confusion and misunderstanding, even among seasoned neutrino aficionados. A coarse scan of the literature reveals several incorrect (but, fortunately, mostly harmless) statements and misunderstandings. For this reason alone, we think such a discussion is useful. Furthermore, a detailed discussion of the physical parameter space for Majorana phases and the physical range for all parameters when there are sterile neutrinos is, to the best of our knowledge, new and, we hope, useful. 

Several subtle aspects of the definition of fermion mixing never surfaced with great prominence in the quark sector, because (a) both quark flavor and mass eigenstates can be produced or detected, and (b) quark mixing angles turn out to be very small. In the lepton sector, even if one ignores the possibility of Majorana neutrinos or sterile neutrino states, the situation is different. For all practical purposes, in the laboratory, all neutrinos are produced as flavor eigenstates, and all neutrinos are detected as flavor eigenstates. Moreover, lepton mixing angles are large (or even maximal) and preempt the most straight forward definition of the different neutrino mass eigenstates.

This paper is organized as follows. In Sec.~\ref{sec:PhyRange}, we carefully define the mixing matrix and describe how unphysical field redefinitions define an equivalence class of physical mixing matrices. We then discuss in detail the two, three and many Majorana neutrino cases. We illustrate some of our results with well-known (and not so well-known) examples. In Sec. \ref{sec:NonStandardInt}, we discuss how the situation would change in the presence of new neutrino interactions. We discuss both lepton-number violating and lepton-number conserving new neutrino interactions. In Sec.~\ref{sec:Conclusion}, we summarize our results and offer some concluding remarks. Three appendices contain, respectively, a detailed discussion of the notation we use to describe unitary matrices of different dimensionality, an example of a ``non-standard'' parameter-space choice that may be useful in the case of Dirac CP-conservation, and approximate expressions for neutrino mixing matrices in the case of three active plus one or two sterile neutrinos. 

\setcounter{equation}{0}
\section{Parameters and Physical Ranges}
\label{sec:PhyRange}

After electroweak symmetry breaking, assuming three sequential generations of leptons that couple to the Standard Model gauge fields, the Standard Model Lagrangian augmented by Majorana neutrino masses, $\mathcal{L}_{\nu SM}$, contains 
\begin{equation}
\mathcal{L}_{\nu SM} \supset 
-\frac{g}{\sqrt{2}}\sum_{\alpha=e,\mu,\tau}
\left(\bar{\nu}_\alpha\gamma^\mu \ell_\alpha W^+_\mu +
\bar{\ell}_\alpha\gamma^\mu\nu_{\alpha} W^-_\mu \right) -
\frac{g}{2\cos\theta_W} \sum_{\alpha=e,\mu,\tau}\bar{\nu}_{\alpha}\gamma^\mu \nu_{\alpha} Z_\mu
- \sum_{\alpha=e,\mu,\tau}m_{\alpha}\bar{e}_{\alpha}\ell_{\alpha} - \frac{1}{2} \sum_i\bar{\nu}_i^c m_i \nu_i.
\label{eq:LSM}
\end{equation}
Here, $\ell_{\alpha}$ ($e_{\alpha}$), $\alpha=e,\mu,\tau$,  are the left-handed (right-handed), charged fermion fields, $\nu_i$, $i=1,2,\ldots,N+3$ are left-handed mass-eigenstate neutrino fields with a well-defined mass, and $\nu_{\alpha}$, $\alpha=e,\mu,\tau,s_1,\ldots,s_N$ are left-handed flavor-eigenstate neutrino fields that couple diagonally to the left-handed charged leptons. $N=0,1,\ldots$ parameterizes the number of sterile neutrinos. Note that for the charged-current and neutral-current neutrino interactions, the sums are restricted only to the so-called active neutrino flavors, $\nu_e,\nu_{\mu},\nu_{\tau}$ \cite{Schechter:1980gr}. 

We choose to work on the weak basis where the charged lepton and neutrino masses ($m_e,m_{\mu},m_{\tau}$ and $m_1,m_2,\ldots, m_{3+N}$, respectively) are real and positive. Neutrino mass eigenstates and flavor eigenstates are related via the neutrino mixing matrix $U$:
\begin{equation}
\nu_{\alpha}=U_{\alpha i}\nu_i,
\label{def_U}
\end{equation}
where $U$ is a unitary $3+N\times 3+N$ matrix: $U_{\alpha i}U^*_{\beta i}=\delta_{\alpha\beta}$, $U_{\alpha i}U^*_{\alpha j}=\delta_{ij}$. The indices $i,j=1,2,\ldots 3+N$ and $\alpha,\beta=e,\mu,\tau,s_1,\ldots,s_N$. We further choose to parameterize the lepton mixing matrix as (see \cite{Haba:2000be,deGouvea:2002gf} for a more detailed discussion)
\begin{equation}
U=\left(\prod_{\alpha=1}^{3+N}\mathbf{P^\alpha}(\phi_\alpha)\right) U'(\theta,\delta)\left(\prod_{i=2}^{3+N}\mathbf{P^i}(\phi_i)\right),
\label{U_par}
\end{equation}
where  $\mathbf{P^i}(\phi)$ is a diagonal $3+N\times 3+N$ matrix whose diagonal entries are all unity except for the $i,i$ element, given by $e^{i\phi}$ (see Appendix~\ref{app:notation}). $U'$ is a unitary matrix that results when one chooses a parameterization such that as many diagonal complex phases are defined to ``to the left'' and ``to the right'' of $U$ as possible.\footnote{If neutrinos were Dirac fermions, $U'$ would contain all of the physically meaningful mixing parameters.} $U'$ is non-diagonal and depends on the henceforth defined mixing angles $\theta$ and Dirac phases $\delta$. Diagonal redefinitions of the $\ell_\alpha$ and $e_{\alpha}$ fields that preserve $m_{\alpha}$ real and positive render all $\phi_{\alpha}$ values physically equivalent.\footnote{There is one subtlety here. In the case of $N\neq0$, there aren't any sterile charged-fermion fields around to field-redefine some of the $\phi_{\alpha}$. Nonetheless, it is easy to see that different values of these phases do not lead to physically distinguishable observables as these parameters do not appear in either the charged-current or neutral-current neutrino interactions and are hence unobservable.} The same is not true of the henceforth defined Majorana phases $\phi_i$ (note that we have already ``spent'' an overall neutrino field redefinition to render the (1,1) element of the matrix on the right-hand side of Eq.~(\ref{U_par}) equal to unity). The only remaining field redefinition that leaves $m_i$ real and positive is to change the sign of $\nu_i$ fields: $\nu_i\to -\nu_i$. Hence, without loss of generality, all $U$ will be parameterized as $U=U'(\theta,\delta)\left(\prod_{i=2}^{3+N}\mathbf{P^i}(\phi_i)\right)$.

We haven't spent all of our ability to redefine neutral and charged lepton fields. From the discussion above, two matrices $U(\theta,\delta,\phi_i)$ and $U(\theta',\delta',\phi'_i)$ describe the same phenomena if 
\begin{equation}
U(\theta',\delta',\phi_i')=\left(\prod_{i=1}^{3+N}\mathbf{P^i}(0,\pi)\right) U(\theta,\delta,\phi_i)\left(\prod_{j=2}^{3+N}\mathbf{P^j}(0,\pi)\right),
\label{master}
\end{equation}
where the diagonal matrices on the left- and right-hand sides have entries that are either 1 or $-1$. Eq.~(\ref{master}) will be used to constrain the physical parameter space of the mixing angles $\theta$, the Dirac phases $\delta$ and Majorana phases $\phi_i$, once the neutrino mass eigenstates are unambiguously defined.

\subsection{The Two Neutrino Case} \label{subsec:TwoNu}

If there were only two active neutrinos (for concreteness, $\nu_e$ and $\nu_{\mu}$), $U$ would be a $2\times 2$ matrix. According to our conventions, $U$ is parameterized by
\begin{equation}
U = \left(
            \begin{array}{cc}
              U_{e1} & U_{e2} \\
              U_{\mu 1} & U_{\mu 2} \\
            \end{array} \right) = \mathbf{R^{12}}(\theta)\mathbf{P^2}(\phi) = 
            \left(
            \begin{array}{cc}
              \cos\theta & \sin\theta \\
              -\sin\theta & \cos\theta \\
            \end{array} \right)
            \left(\begin{array}{cc}
              1 & 0 \\
              0 & e^{i\phi} \\
            \end{array}\right)
         \label{eq:TwoNuMix},
\end{equation}
where the matrix $\mathbf{R^{12}}$ is defined in Appendix~\ref{app:notation}.

The physical parameter space for $\theta$ and $\phi$ depends on the definition of neutrino mass eigenstates and neutrino flavor eigenstates (or weak eigenstates). The definition of neutrino flavor eigenstates is unambiguous: the electron neutrino $\nu_e$ is the chiral field that couples to the electron and the $W$-boson fields, while the electron is the lightest charged-fermion. The same goes for $\nu_{\mu}$. It is the neutrino that couples to the muon, which in turn is the heaviest charged-lepton. The neutrino mass eigenstates $\nu_1$ and $\nu_2$ need to be similarly defined. Here, we discuss two different definitions. Note that both uniquely specify the neutrino mass eigenstates.
\begin{enumerate}
\item[A2.] $\nu_1$ is the lightest state, while $\nu_2$ is the heaviest state: $m_1<m_2$.
\item[B2.] $\nu_1$ is the state with the largest $\nu_e$ content, while $\nu_2$ is the state with the largest $\nu_{\mu}$ content. The $\nu_{\alpha}$ content of $\nu_i$ is define to be $|U_{\alpha i}|^2$: $|U_{e1}|^2>|U_{e2}|^2$.
\end{enumerate}
Direct examination of Eq.~(\ref{eq:TwoNuMix}) reveals that, at most, $\theta\in[-\pi,\pi[$ and $\phi\in[-\pi,\pi[$ (case A2) and $\theta\in[-\pi,-3\pi/4]\cup [-\pi/4,\pi/4] \cup [3\pi/4,\pi[ $ and $\phi\in[-\pi,\pi[$ (case B2).

Eq.~(\ref{master}) is now used to further constrain the  physical ranges of $\theta$ and $\phi$. This is done by identifying, given a fixed value of $\theta$ and $\phi$, all $\theta'$ and $\phi'$ that satisfy
\begin{eqnarray}
{\bf R^{12}}(\theta^\prime) {\bf P^2}(\phi^\prime) & = & {\bf N_f}
{\bf R^{12}}(\theta){\bf P^2}(\phi) {\bf N_m},~~~{\rm or}\\
{\bf R^{12}}(\theta^\prime) {\bf P^2}(\phi^\prime)\mathbf{N_{m}} & = & {\bf N_f}
{\bf R^{12}}(\theta){\bf P^2}(\phi).
\label{eq:2NuCondition}
\end{eqnarray}
where $\mathbf{N_f}=\mathbf{I},{\bf P^1}(\pi),{\bf P^2}(\pi),-\mathbf{I}$ and $\mathbf{N_m}=\mathbf{I},{\bf P^2}(\pi)$, where $\mathbf{I}$ is the identify matrix ($2\times 2$ in this case). The results of all these equations (some of which are redundant) are easy to obtain using the algebra presented in Appendix~\ref{app:notation}, and are tabulated in Table~\ref{tab:2nus}. For example,
\begin{eqnarray}
-\mathbf{I}\mathbf{R^{12}}(\theta)\mathbf{P^2}(\phi)&=&\mathbf{R^{12}}(\theta+\pi)\mathbf{P^2}(\phi)=\mathbf{R^{12}}(\theta+\pi)\mathbf{P^2}(\phi)\mathbf{I},
\end{eqnarray}
so that $\mathbf{N_f}=-\mathbf{I}$ and $\mathbf{N_m}=\mathbf{I}$ lead to $\theta'=\theta+\pi$ and $\phi'=\phi$.
\begin{table}
\caption{Values of $\theta'$ and $\phi'$ that satisfy Eq.~(\ref{eq:2NuCondition}), for different values of $\mathbf{N_f}$ and $\mathbf{N_m}$ (see text for details). Also tabulated are the field redefinitions corresponding to the different choices of $\mathbf{N_f}$ and $\mathbf{N_m}$.}
\begin{tabular}{|c|c|c|}
  \hline
  $\mathbf N_f$, $\mathbf N_m$ & Field Redefinition & $\theta'$, $\phi'$ \\
 \hline
    $\mathbf{I}$, $\mathbf{I}$ & none & $\theta$, $\phi$ \\
    $\mathbf{I}$, $\mathbf{P^2}(\pi)$ & $\nu_2\to-\nu_2$ & $\theta$, $\phi+\pi$ \\
    $\mathbf{P^2}(\pi)$, $\mathbf{I}$ & $\ell_{\mu}\to -\ell_{\mu}$ & $-\theta$, $\phi+\pi$ \\
    $\mathbf{P^2}(\pi)$, $\mathbf{P^2}(\pi)$ & $\ell_{\mu}\to-\ell_{\mu}$, $\nu_2\to-\nu_2$ & $-\theta$, $\phi$ \\
    $\mathbf{-I}$, $\mathbf{I}$ & $\ell_{e}\to -\ell_{e}$, $\ell_{\mu}\to -\ell_{\mu}$ & $\theta+\pi$, $\phi$ \\
    $\mathbf{-I}$, $\mathbf{P^2}(\pi)$ & $\ell_{e}\to -\ell_{e}$, $\ell_{\mu}\to -\ell_{\mu},\nu_2\to-\nu_2$ & $\theta+\pi$, $-\phi$ \\
  \hline
\end{tabular}
\label{tab:2nus}
\end{table}

The table (and the example above) reveals that $\theta$ and $\theta+\pi$ are equivalent. Hence, it is enough to restrict oneself to $\theta\in [-\pi/2,\pi/2]$ (case A2) or $[-\pi/4,\pi/4]$ (case B2). Furthermore, $\theta$ and $-\theta$ are also equivalent. This allows one to further restrict $\theta\in [0,\pi/2]$ (case A2) or $[0,\pi/4]$ (case B2). Finally, the Majorana phase is independently invariant under $\phi\to\phi+\pi$. It can therefore be constrained to $\phi\in [0,\pi]$ (cases A2 and B2). This result agrees with a recent  analysis of leptonic rephasing invariants \cite{Jenkins:2007ip}. 

The reason for the different physical parameter space for $\theta$ in case A2 and B2 has nothing to do with the redundancies of the fields and parameters, but is a consequence of our definition of the mass eigenstates. It is clear that both choices describe all different physics scenarios equivalently well. This is accomplished by noting that, in case B2, one must allow for {\sl both} $m_1<m_2$ {\sl and} $m_1>m_2$. This is not true in case A2. It possible to map case A2 into case B2. One can write
\begin{eqnarray} 
&U(\theta,\phi)=\mathbf{P^2}(\pi)\mathbf{R^{12}}(\pi/2-\theta)\mathbf{P^2}(-\phi)\mathbf{R^{12}}(\pi/2)e^{i\phi}, \\ {\rm or} \nonumber \\
&U(\theta,\phi)= \left(
            \begin{array}{cc}
              e^{i\phi} & 0 \\
              0 & -e^{i\phi} \\
            \end{array} \right)
            \left(
            \begin{array}{cc}
              \cos(\pi/2-\theta) & \sin(\pi/2-\theta) \\
              -\sin(\pi/2-\theta) & \cos(\pi/2-\theta) \\
            \end{array} \right)
            \left(
            \begin{array}{cc}
              1 & 0 \\
              0 & e^{-i\phi} \\
            \end{array} \right)
            \left(
            \begin{array}{cc}
              0 & 1 \\
              1 & 0 \\
            \end{array} \right).
\end{eqnarray}
Hence, a mixing matrix with $0\le\theta\le\pi/4$ can be mapped into another mixing matrix with  $\pi/4\le\theta\le\pi/2$ if the sign of $\phi$ is reversed and if one ``exchanges'' $\nu_1\leftrightarrow\nu_2$ (up to an allowed redefinition of the left-handed lepton fields). This means that  points  $(\theta,m_1,m_2,\phi)$  in the case A2 parameter space associated to $\pi/4<\theta\le\pi/2$ are equivalent to the points $(\pi/2-\theta,m_2,m_1,-\phi)$ in the case B2 parameter space. This will be concretely illustrated in the examples below.

The probability $P_{e\mu}^{\rm vac}$ of a neutrino produced as a $\nu_e$ with energy $E$ to be detected as $\nu_{\mu}$ after propagating a distance $L$ in vacuum is 
\begin{equation}
P_{e\mu}^{\rm vac} = \sin^2
2\theta \sin^2 \left( \frac{\Delta m^2 L}{4E} \right),
\label{eq:VacNuOss}
\end{equation}
where $\Delta m^2\equiv m_2^2-m_1^2$. In case A2, $\Delta m^2$ is positive-definite, while in case B2 it can be either positive or negative. It is clear that this observable does not depend on the Majorana phase $\phi$. It is also clear that $\theta$ and $-\theta$ and $\pi+\theta$ lead to the same oscillation probability. Furthermore, in case A2, $\theta$ and $\pi/2-\theta$ also yield the same value of $P_{e\mu}^{\rm vac}$ for fixed $\Delta m^2$. This implies that in case B2, for fixed $\theta$, $\Delta m^2$ and $-\Delta m^2$ should also yield the same value of $P_{e\mu}^{\rm vac}$. This is, of course, the case.

The probability $P_{e\mu}^{\rm matter}$ of a neutrino produced as a $\nu_e$ with energy $E$ to be detected as $\nu_{\mu}$ after propagating a distance $L$ in a medium characterized by a constant electron background is
\begin{equation}
P_{e\mu}^{\rm matter} = \frac{\sin^2
2\theta}{1+A^2-2A\cos 2\theta} \sin^2 \left( \frac{\Delta m^2
L}{4E}\sqrt{1+A^2-2A\cos 2\theta} \right), \label{eq:MatNuOss}
\end{equation}
where $A = 2\sqrt{2}E G_F N_e/\Delta m^2$, $G_F$ is the Fermi constant, and $N_e$ is the local electron number density. As before, $\Delta m^2\equiv m_2^2-m_1^2$. As in the case of Eq.~(\ref{eq:VacNuOss}), the Majorana phase $\phi$ plays no role in determining $P_{e\mu}^{\rm matter}$, and one can explicitly see that $\theta$ and $-\theta$ and $\theta$ and $\pi+\theta$ lead to the same value of $P_{e\mu}^{\rm matter}$. In case A2, $\theta$ and $\pi/2-\theta$ lead to different values of 
$P_{e\mu}^{\rm matter}$ for fixed $\Delta m^2$ since $\theta\to\pi/2-\theta$ leads to $\cos2\theta\to -\cos2\theta$ and $P_{e\mu}^{\rm matter}$ depends on the sign of the $\cos2\theta$ term \cite{de Gouvea:2000cq}. In case B2, one can see that $\Delta m^2$ and $-\Delta m^2$ also lead to different values for $P_{e\mu}^{\rm matter}$ for fixed $\theta$. Furthermore, as advertised earlier, the points $(\theta,\Delta m^2)$ and $(\pi/2-\theta,\Delta m^2)$ in case A2 are equivalent to the points $(\theta,\Delta m^2)$ and $(\theta,-\Delta m^2)$ in case B2.

The value of the Majorana phase will only affect processes where lepton number is violated. As an example, consider the rate for neutrino--antineutrino oscillations, related to the probability that a neutrino produced as a $\nu_{e}$ with energy $E$ is detected, after propagating some distance $L$, as what appears to be a $\bar{\nu}_{\mu}$. It is \cite{deGouvea:2002gf,nu_nubar}
\begin{equation}
P_{e\bar{\mu}}\propto\frac{\sin^2 2\theta}{4E^2}\left\{m_1^2 + m_2^2 - 2m_1 m_2 \cos
\left(\frac{\Delta m^2 L}{2E} - 2\phi \right) \right\}.
\end{equation}
As in the previous cases, it is clear that $\theta\in [0,\pi/2]$ covers all distinct values of this neutrino--antineutrino oscillation probability since $P_{e\bar{\mu}}$ is only a function of $\sin^22\theta$. One can also see that $\phi$ and $\pi+\phi$ lead to the same expression. In the case A2 parameterization, one sees that for fixed $m_1$ and $m_2$, $\theta$ and $\pi/2-\theta$ lead to the same expression. This is also true using parameterization B2 and exchanging $m_1\leftrightarrow m_2$ {\sl and}  $\phi\to -\phi$. To the best of our knowledge, this ``invariance'' was first noticed in \cite{deGouvea:2002gf}.

\subsection{The Three Neutrino Case} \label{subsec:ThreeNu}

In the case of three active neutrinos and no sterile neutrinos ($N=0$) $U$ is a $3\times 3$ matrix. According to our conventions, $U$ is parameterized by 
\begin{eqnarray}
U&=& {\bf
R^{23}}(\theta_{23})
{\bf P^3}(-\delta/2){\bf P^1}(\delta/2){\bf
R^{13}}(\theta_{13}){\bf P^3}(\delta/2){\bf P^1}(-\delta/2) {\bf R^{12}}(\theta_{12}){\bf P^2}(\phi_2){\bf P^3}(\phi_3),   \\
&&\nonumber \\
 &=&  
 \left(
            \begin{array}{ccc}
              1 & 0 & 0 \\
              0 & c_{23} & s_{23} \\
              0 & -s_{23} & c_{23} \\
            \end{array}
          \right)
          \left(
            \begin{array}{ccc}
              c_{13} & 0 & s_{13}e^{i\delta} \\
              0 & 1 & 0 \\
              -s_{13}e^{-i\delta} & 0 & c_{13} \\
            \end{array}
          \right)
          \left(
            \begin{array}{ccc}
              c_{12} & s_{12} & 0 \\
              -s_{12} & c_{12} & 0 \\
              0 & 0 & 1 \\
            \end{array}
          \right)
          \left(
            \begin{array}{ccc}
              1 & 0 & 0\\
              0 & e^{i\phi_2} & 0 \\
              0 & 0 & e^{i\phi_3} \\
            \end{array}
          \right)
 \label{eq:ThreeNuMix},
\end{eqnarray}
where, to preserve space, we take advantage of the shorthand notation $c_{ij}\equiv \cos
\theta_{ij}$ and $s_{ij} \equiv \sin \theta_{ij}$, $ij=12,13,23$.

As in the previous subsection, the question of the physical parameter space for the mixing angles $\theta_{12},\theta_{13},\theta_{23}$ and the CP-violating phases $\delta,\phi_2,\phi_3$ depends on the proper definition of the neutrino mass eigenstates, $\nu_1$, $\nu_2$, and $\nu_3$. Here we discuss three cases:
\begin{enumerate}
\item[A3.] $\nu_1$ is the lightest state, $\nu_2$ is the second lightest state, $\nu_3$ is the heaviest state: $m_1<m_2<m_3$.
\item[B3.] $\nu_1$ is the state with the largest $\nu_e$ content, $\nu_2$ is the state with the second largest $\nu_e$ content, and $\nu_3$ is the state with the smallest $\nu_e$ content: $|U_{e1}|^2>|U_{e2}|^2>|U_{e3}|^2$.
\item[C3.] The $\nu_1$ and $\nu_2$ states are defined such that $\nu_1$ is always lighter than $\nu_2$ and $m_2^2-m_1^2$ is, in magnitude, the smallest mass-squared difference. In this case, $m_1<m_2<m_3$ if $\Delta m^2_{13}>0$ and $m_3<m_1<m_2$ if $\Delta m^2_{13}<0$. The former is referred to as the `normal' mass hierarchy, while the latter is the `inverted' mass hierarchy. Note that we define $\Delta m^2_{ij}=m_j^2-m_i^2$.
\end{enumerate}

Direct examination of Eq.~(\ref{eq:ThreeNuMix}) reveals that, at most, $\theta_{ij}\in[-\pi,\pi[$ ($ij=12,13,23$) and $\phi_2,\phi_3,\delta\in[-\pi,\pi[$ (cases A3 and C3). For case B3, the situation is more complicated. Naively, $\theta_{23}\in[-\pi,\pi[$ ($ij=12,13,23$) and $\phi_2,\phi_3,\delta\in[-\pi,\pi[$, while $\sin^2\theta_{12}\le 1/2$ and $\tan^2\theta_{13}\le\sin^2\theta_{12}$. 
Furthermore, since $\sin\theta_{13}$ and $\delta$ always appear as the combination $\sin\theta_{13}e^{\pm i\delta}$ we are allowed to either constrain $\sin\theta_{13}>0$ or $\delta>\in[0,\pi]$. One can also check that Eq.~(\ref{eq:ThreeNuMix}) is unchanged if one simultaneously redefine all mixing angles $\theta_{ij}\to\theta_{ij}+\pi$, and $\delta\to\delta+\pi$. Both of these redundancies are included in Table~\ref{tab:3nus} (see below).

In order to further constrain the mixing parameter space, we search for all $\theta'_{ij},\delta',\phi'_2,\phi'_3$ ($ij=12,13,23$) that satisfy 
\begin{eqnarray}
U(\theta'_{ij},\phi'_2,\phi'_3,\delta') & = & \mathbf{N_f}U(\theta_{ij},\phi_2,\phi_3,\delta)\mathbf{N_m},~~~{\rm or} \\
U(\theta'_{ij},\phi'_2,\phi'_3,\delta')\mathbf{N_m} & = & \mathbf{N_f}U(\theta_{ij},\phi_2,\phi_3,\delta).
\label{eq:3NuCondition}
\end{eqnarray}
Here, $\mathbf{N_f}$ is equal to $\mathbf{I}$, $\mathbf{P^i}(\pi)$ ($i=1,2,3$), $\mathbf{R^{ij}}(\pi)\equiv\mathbf{P^i}(\pi)\mathbf{P^j}(\pi)$ ($ij=12,13,23$), or $-\mathbf{I}=\mathbf{P^1}(\pi)\mathbf{P^2}(\pi)\mathbf{P^3}(\pi)$. On the other hand, $\mathbf{N_m}$ is equal to $\mathbf{I}$, $\mathbf{P^i}(\pi)$ ($i=2,3$), or $\mathbf{R^{23}}(\pi)=\mathbf{P^2}(\pi)\mathbf{P^3}(\pi)$. Before proceeding, we need to qualify the types of solutions we are looking for. We are only interested in linear solutions of the type $\vartheta'=\vartheta+k$, where $k$ is constant and $\vartheta$ is a generic mixing parameter. These solutions allow one to relate disconnected regions of the parameter space and can be used to reduce the volume of the original space. We do not rule out the possibility that other solutions exists, but these will only apply to ``lower-dimensional'' subspaces of the parameter space and are not useful when it comes to defining the physically distinguisable values of the mixing parameters. 

For example, 
\begin{eqnarray}
U(\theta_{ij},\delta,\phi_2,\phi_3)\mathbf{R^{23}}(\pi) & =
& \left[{\bf
R^{23}}(\theta_{23})
{\bf P^3}(-\delta/2){\bf P^1}(\delta/2){\bf
R^{13}}(\theta_{13}){\bf P^3}(\delta/2){\bf P^1}(-\delta/2) {\bf R^{12}}(\theta_{12}){\bf P^2}(\phi_2){\bf P^3}(\phi_3)\right]\mathbf{R^{23}}(\pi), \nonumber \\
& = & \left[{\bf
R^{13}}(\theta_{13}){\bf P^3}(\delta/2){\bf P^1}(-\delta/2) {\bf R^{12}}(\theta_{12}){\bf P^2}(\phi_2+\pi){\bf P^3}(\phi_3+\pi)\right] \label{sample1}, \\
& =
&  \left[{\bf
R^{23}}(\theta_{23}+\pi)
{\bf P^3}(-\delta/2){\bf P^1}(\delta/2){\bf
R^{13}}(-\theta_{13}){\bf P^3}(\delta/2){\bf P^1}(-\delta/2) {\bf R^{12}}(-\theta_{12}){\bf P^2}(\phi_2){\bf P^3}(\phi_3)\right], \nonumber \\ \label{sample2}
\end{eqnarray}
where we made use of the notation and the product rules described in Appendix~\ref{app:notation}. Eqs.~(\ref{sample1},\ref{sample2}) indicate that the mixing matrices parameterized by $\{\theta_{12},\theta_{13},\theta_{23},\delta,\phi_2,\phi_3\}$, $\{\theta_{12},\theta_{13},\theta_{23},\delta,\phi_2+\pi,\phi_3+\pi\}$, and $\{-\theta_{12},-\theta_{13},\theta_{23}+\pi,\delta,\phi_2,\phi_3\}$ describe the same physics. Many other solutions for $\theta'_{ij},\delta',\phi'_2,\phi'_3$ are tabulated in Table~\ref{tab:3nus}, including all of the independent ones. Several more, none of which provide any extra constraints, have been omitted. 
\begin{table}
\caption{Values of $\theta'_{12,13,23}$ and $\delta,\phi'_2,\phi_3$ that satisfy Eq.~(\ref{eq:3NuCondition}), for different values of $\mathbf{N_f}$ and $\mathbf{N_m}$ (see text for details). Also tabulated are the field redefinitions corresponding to the different choices of $\mathbf{N_f}$ and $\mathbf{N_m}$.}
\begin{tabular}{|c|c|c|c|}
  \hline
  $\mathbf N_f$, $\mathbf N_m$ & Field Redefinition & $\theta'_{12},\theta'_{13},\theta'_{23}$ & $\delta,\phi'_2,\phi'_3$ \\
 \hline
    $\mathbf{I}$, $\mathbf{I}$ & none & $\theta_{12},\theta_{13},\theta_{23}$ & $\delta,\phi_2,\phi_3$ \\
     $\mathbf{I}$, $\mathbf{I}$ & none & $\theta_{12},-\theta_{13},\theta_{23}$ & $\delta+\pi,\phi_2,\phi_3$ \\
      $\mathbf{I}$, $\mathbf{I}$ & none & $\theta_{12}+\pi,\theta_{13}+\pi,\theta_{23}+\pi$ & $\delta+\pi,\phi_2,\phi_3$ \\
    $\mathbf{I}$, $\mathbf{P^2}(\pi)$ & $\nu_2\to-\nu_2$ & $\theta_{12},\theta_{13},\theta_{23}$ & $\delta,\phi_2+\pi,\phi_3$\\
     $\mathbf{I}$, $\mathbf{P^3}(\pi)$ & $\nu_3\to-\nu_3$ & $\theta_{12},\theta_{13},\theta_{23}$ & $\delta,\phi_2,\phi_3+\pi$\\
    $\mathbf{I}$, $\mathbf{R^{23}}(\pi)$ & $\nu_2\to-\nu_2$, $\nu_3\to-\nu_3$ & $\theta_{12},\theta_{13},\theta_{23}$ & $\delta,\phi_2+\pi,\phi_3+\pi$\\
     $\mathbf{I}$, $\mathbf{R^{23}}(\pi)$ & $\nu_2\to-\nu_2,\nu_3\to-\nu_3$ & $-\theta_{12},-\theta_{13},\theta_{23}+\pi$ & $\delta,\phi_2,\phi_3$\\
     $\mathbf{-I}$, $\mathbf{P^{3}}(\pi)$ & $\ell_{\alpha}\to-\ell_{\alpha}$ ($\alpha=e,\mu,\tau$), $\nu_3\to-\nu_3$ & $\theta_{12}+\pi,\theta_{13},\theta_{23}$ & $\delta,\phi_2,\phi_3$ \\
     $\mathbf{-I}$, $\mathbf{P^{2}}(\pi)$ & $\ell_{\alpha}\to-\ell_{\alpha}$ ($\alpha=e,\mu,\tau$), $\nu_2\to-\nu_2$ & $-\theta_{12},\theta_{13}+\pi,\theta_{23}$ & $\delta,\phi_2,\phi_3$ \\
$\mathbf{P^2}(\pi)$, $\mathbf{P^2}(\pi)$ & $\ell_{\mu}\to -\ell_{\mu}$, $\nu_2\to-\nu_2$ & $-\theta_{12},\theta_{13},-\theta_{23}$ & $\delta,\phi_2,\phi_3$\\
$\mathbf{P^3}(\pi)$, $\mathbf{P^3}(\pi)$ & $\ell_{\tau}\to -\ell_{\tau}$, $\nu_3\to-\nu_3$ & $\theta_{12},-\theta_{13},-\theta_{23}$ & $\delta,\phi_2,\phi_3$\\
$\mathbf{R^{12}}(\pi)$, $\mathbf{I}$ & $\ell_{e}\to -\ell_{e},\ell_{\mu}\to -\ell_{\mu}$ & $\theta_{12}+\pi,-\theta_{13},-\theta_{23}$ & $\delta,\phi_2,\phi_3$\\
$\mathbf{R^{13}}(\pi)$, $\mathbf{I}$ & $\ell_{e}\to -\ell_{e},\ell_{\tau}\to -\ell_{\tau}$ & $\theta_{12},\theta_{13}+\pi,-\theta_{23}$ & $\delta,\phi_2,\phi_3$\\
$\mathbf{R^{23}}(\pi)$, $\mathbf{I}$ & $\ell_{\mu}\to -\ell_{\mu},\ell_{\tau}\to -\ell_{\tau}$ & $\theta_{12},\theta_{13},\theta_{23}+\pi$ & $\delta,\phi_2,\phi_3$\\
$\mathbf{R^{23}}(\pi)$, $\mathbf{R^{23}}(\pi)$ & $\ell_{\mu}\to -\ell_{\mu},\ell_{\tau}\to -\ell_{\tau}$, $\nu_2\to-\nu_2,\nu_3\to-\nu_3$ & $-\theta_{12},-\theta_{13},\theta_{23}$ & $\delta,\phi_2,\phi_3$\\
 \hline
\end{tabular}
\label{tab:3nus}
\end{table}

Table~\ref{tab:3nus} reveals that for fixed mixing angles and $\delta$, matrices with $\phi_2$ and $\phi_2+\pi$ are physically indistinguishable. This also holds for $\phi_3$ and $\phi_3+\pi$. Hence both can be constrained: $\phi_2,\phi_3\in[0,\pi]$ (all cases A3, B3, and C3). Keeping all parameters fixed, one can easily show that matrices with $\theta_{12}$ and $\theta_{12}+\pi$ are identical, the same holding for $\theta_{23}$ and $\theta_{23}+\pi$ and for $\theta_{13}$ and $-(\theta_{13}+\pi)$.
This allows one to further constrain the ranges for the mixing angles to $\theta_{ij}\in[-\pi/2,\pi/2[$ ($ij=12,13,23$) (cases A3 and C3) and $\theta_{12}\in[-\pi/4,\pi/4]$, $\theta_{13}\in[-\arctan(|\sin\theta_{12}|),\arctan(|\sin\theta_{12}|)]$, $\theta_{23}\in[-\pi/2,\pi/2[$ (case B3). Finally, points with $\theta_{ij},\delta$ and $-\theta_{ij},\delta +\pi$ (for any particular $ij=12,13,23$) are also physically indistinguishable. We will then choose all mixing angles positive, which implies that the physical range for the Dirac phase is $\delta\in[-\pi,\pi[$. We are left with
\begin{itemize}
\item cases A3 and C3: $\theta_{12},\theta_{13},\theta_{23}\in[0,\pi/2]$, $\delta\in[-\pi,\pi[$, $\phi_2,\phi_3\in [0,\pi]$;
\item case B3:  $\theta_{12}\in[0,\pi/4]$,  $\theta_{13}\in[0,\arctan(\sin\theta_{12})]$, $\theta_{23}\in[0,\pi/2]$, $\delta\in[-\pi,\pi[$, $\phi_2,\phi_3\in [0,\pi]$.
\end{itemize}
This choice is not unique, and is a consequence of allowing $\delta$ to take all of its naively allowed values. It also agrees with the choice made in virtually all of the neutrino literature and seems to be the most natural one. We find one circumstance where a different choice may be advantageous from a pragmatic point of view. Using the parameterization above, the case of Dirac CP-conservation is contained in two ``disjoint'' subsets of the parameter space: $\delta=0$ and $\delta=\pi$.  On the other hand, if one chose $\theta_{13}$ such that negative and positive values were included, $\delta$ could be constrained to lie within $[-\pi/2,\pi/2]$. In this case, the Dirac CP-conserving subspace is continuous and, {\it e.g.}, in case C3, $\theta_{13}$ would lie within $[-\pi/2,\pi/2]$. This observation was first made by the authors of \cite{Latimer:2004hd}. A concrete numerical example of this is depicted in Appendix~\ref{neg_ue3}.

As in the two-neutrino case, one can map one definition of neutrino mass eigenstates into another. Case C3 turns out to be the one that is widely utilized in the literature \cite{NeutrinoReview}, with case A3 a distant second \cite{Petcov:2001sy}.\footnote{Case B3 was inspired by a comment made by Alexei Smirnov to one the authors. To the best of our knowledge, case B3 has not been explicitly considered in the literature. We can't, however, rule out the possibility that it has been implicitly assumed in several occasions.} Unlike the two neutrino case, the map between different cases is more involved. Here, for illustrative purposes, we discuss briefly how one can relate case A3 and case C3.

\begin{figure}[t]
\begin{center}
\includegraphics[scale=.50]{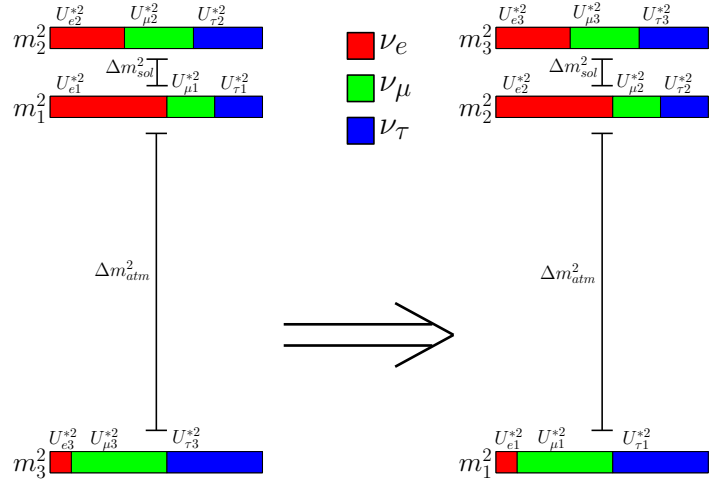}
\caption{Identical copies of a particular point in the lepton parameter space (charged-fermion masses not included), one labeled according to the conventions of case A3 (right-hand side), the other according to the conventions of case C3 (left-hand side). See text for details.}\label{fig:InvMap}
\end{center}
\end{figure}

If the neutrino mass hierarchy is normal (in case C3 language, $\Delta m^2_{13}>0$), case A3 and case C3 are identical. If the mass hierarchy is inverted (in case C3 language, $\Delta m^2_{13}<0$), however, the situation is more interesting. Fig.~\ref{fig:InvMap} depicts two identical copies of a particular point in the lepton parameter space (charged-fermion masses not included), one labeled according to the conventions of case A3 (right-hand side), the other according to the conventions of case C3 (left-hand side).  In order to convert from case C3~$\to$~case A3, one need only ``relabel'' the mass eigenstates as follows: $3\to 1$,  $1\to 2$, $2\to 3$. 

Such a relabeling will be accompanied by a non-trivial mapping of mixing angles. Fig.~\ref{fig:InvMap} reveals, for example, that $U^{(312)}_{e3} = U^{(123)}_{e1}$, which implies that $s_{13}^{(312)}e^{i\delta^{(312)}} = c_{12}^{(123)}c_{13}^{(123)}$, where the parameter superscripts indicate the mass labeling (312 is an inverted mass ordering in case C3, 123 is the same picture in case A3).  We proceed, using the language of Appendix~\ref{app:notation}, to formally complete the mapping.  Specifically, we study the effect of 
\begin{equation}
{\bf H}\left(
  \begin{array}{c}
    \nu_1 \\
    \nu_2 \\
    \nu_3 \\
  \end{array}
\right) = {\bf
R^{13}}\left(\frac{\pi}{2}\right){\bf R^{23}}\left(\frac{\pi}{2}\right)\mathbf{R^{12}}(\pi)\left(
  \begin{array}{c}
    \nu_1 \\
    \nu_2 \\
    \nu_3 \\
  \end{array}
\right) = \left(
  \begin{array}{c}
    \nu_2 \\
    \nu_3 \\
    \nu_1 \\
  \end{array}
\right)
\end{equation}
on the mixing matrix. Acting on the mixing matrix, this has the following effect:
\begin{eqnarray}
U\mathbf{H^{-1}}&=& {\bf
R^{23}}(\theta_{23})
{\bf P^3}\left(-\frac{\delta}{2}\right){\bf P^1}\left(\frac{\delta}{2}\right){\bf
R^{13}}(\theta_{13}){\bf P^3}\left(\frac{\delta}{2}\right){\bf P^1}\left(-\frac{\delta}{2}\right) {\bf R^{12}}(\theta_{12}){\bf P^2}(\phi_2){\bf P^3}(\phi_3)\mathbf{R^{12}}(\pi){\bf R^{23}}\left(-\frac{\pi}{2}\right) {\bf
R^{13}}\left(-\frac{\pi}{2}\right), \nonumber \\
&=& {\bf
R^{23}}\left(\theta_{23}-\frac{\pi}{2}\right)
{\bf P^2}\left(\frac{-\delta}{2}\right){\bf P^1}\left(\frac{\delta}{2}\right){\bf
R^{12}}(\theta_{13}){\bf P^2}\left(\frac{\delta}{2}\right){\bf P^1}\left(\frac{-\delta}{2}\right) {\bf R^{13}}(\theta_{12}+\pi){\bf P^3}(\phi_2){\bf P^2}(\phi_3){\bf R^{13}}\left(-\frac{\pi}{2}\right), \nonumber \\
&=& e^{i\phi_2} {\bf
R^{23}}\left(\theta_{23}-\frac{\pi}{2}\right)
{\bf P^2}\left(\frac{-\delta}{2}\right){\bf P^1}\left(\frac{\delta}{2}\right){\bf
R^{12}}(\theta_{13}){\bf P^2}\left(\frac{\delta}{2}\right){\bf P^1}\left(\frac{-\delta}{2}\right) {\bf R^{13}}\left(-\theta_{12}+\frac{\pi}{2}\right){\bf P^3}(-\phi_2){\bf P^2}(\phi_3-\phi_2).  \nonumber \\
\label{eq:InvFormalMap}
\end{eqnarray}
The entire transformation may be absorbed into shifts of
the physical mixing parameters (plus re-identifications of the Majorana phases) up to an overall field redefinition of all charged-lepton fields.  
Formally, the transformations shown in
Eqs.~(\ref{eq:InvFormalMap}) are very simple and can be summarized as a relabeling of the mixing planes. This leads to a
situation where the complex rotation governed by $\theta_{13}$ and
$\delta$ now acts in the $1-2$ plane, while $\theta_{12}$ governs
the $1-3$ rotation ---  the ``roles'' of $\theta_{12}$ and $\theta_{13}$ are exchanged.   
To express the result of Eq.~(\ref{eq:InvFormalMap}) in the form of Eq.~(\ref{eq:ThreeNuMix}), however, one must ``commute'' the $1-2$ and $1-3$ rotations and properly redefine the $\theta_{23}$ mixing angle.  We find the following mapping of mixing parameters from the $312$ (Fig.~\ref{fig:InvMap}(left), case C3) to the $123$ (Fig.~\ref{fig:InvMap}(right), case A3) pictures.
\begin{equation}
\begin{array}{lll}
{\bf (312)} &\rightarrow& {\bf (123)}\\
\Delta m_{13}^2 & \rightarrow & -\Delta m_{12}^2 \\
\Delta m_{23}^2 & \rightarrow & -\Delta m_{31}^2 \\
\Delta m_{12}^2 & \rightarrow & \Delta m_{23}^2 \\
\end{array} \Rightarrow \left\{
\begin{array}{lll}
C_{12}^2 &=& \frac{s_{13}^2}{1-c_{13}^2s_{12}^2},\\
S_{13}^2 &=& c_{13}^2s_{12}^2,\\
C_{23}^2 &=& \frac{c_{23}^2s_{12}^2s_{13}^2 + c_{12}^2s_{23}^2 + 2\cos\delta c_{12}c_{23}s_{12}s_{23}s_{13}}{1 - c_{13}^2s_{12}^2},\\
\Delta &=& \cos^{-1}\left( \frac{S_{12}^2S_{23}^2S_{13}^2 +
C_{23}^2\left(C_{12}^2 - C_{13}^2 \right) + c_{13}^2\left(c_{12}^2 -
c_{23}^2 \right)}{2C_{12}S_{12}C_{23}S_{23}S_{13}}
\right),\\
\Phi_3 &=& \phi_2 - \phi_3,  \\
\Phi_2 &=& -\phi_3, \\
\end{array}
\right.
\end{equation}
where capital-letter parameters ($C_{12},\Delta,\Phi_3$, etc) stand for parameters in the 312 picture (Fig.~\ref{fig:InvMap}(left), case C3) while lower-case letter parameters ($c_{12},\delta,\phi_3$, etc) stand for parameters in the 123 picture (Fig.~\ref{fig:InvMap}(right), case A3).
These transformations are unique up to trivial sign redefinitions, provided the overall
nonphysical rephasing by $e^{i(\delta + \phi_3)}$.  The
transformations are relatively simple for the $\theta_{12}$ and
$\theta_{13}$ parameters, but the redefinition is less attractive
for $\theta_{23}$ due to its induced dependence on $\delta$.  In the
case of ``Dirac'' CP conservation --- $\delta=0$ and $\delta=\pi$ ---
the value of $\delta$ is independent of the definition of the mass eigenstates (as expected), and the
more cumbersome mappings are significantly simplified, {\it e.g.}
$c_{23}^2 \rightarrow \left(c_{23}s_{12}s_{13} \pm c_{12}s_{23}
\right)^2/\left(1-c_{13}^2s_{12}^2\right)$, $\delta\to\delta+\pi$.

It is useful to discuss a couple of well-known examples. In vacuum, the probability that a neutrino produced as a $\nu_{\alpha}$ will be detected as a $\nu_{\beta}$ is
\begin{eqnarray}
P_{\alpha\beta}&=&\left|\sum_i U_{\alpha i}U^*_{\beta i}e^{i\Delta m^2_{1i}L/2E}\right|^2, \\
&=&\delta_{\alpha\beta}-4\sum_{ij=12,}^{13,23}\sin^2\left(\frac{\Delta m^2_{ij}L}{4E}\right)\Re\left[U_{\alpha i}U_{\alpha j}^*U_{\beta i}^*U_{\beta j}\right] + 2\sum_{ij=12,}^{13,23}\sin\left(\frac{\Delta m^2_{ij}L}{2E}\right)\Im\left[U_{\alpha i}U_{\alpha j}^*U_{\beta i}^*U_{\beta j}\right] . 
\label{Pvac3}
\end{eqnarray}
It is simple, if not tedious, to check that all terms in the $ij$ sum in Eq.~(\ref{Pvac3}) are invariant under all transformations in 
Table~\ref{tab:3nus} \cite{Gluza:2001de}. For example, $\alpha=\mu$, $\beta=\tau$, $ij=13$ leads to 
\begin{eqnarray}
\Re\left[U_{\mu 1}U_{\mu 3}^*U_{\tau 1}^*U_{\tau 3}\right]&=&\frac{1}{4}\left[s^22\theta_{23}\left(\frac{1}{4}s^22\theta_{13}c^2\theta_{12}-c^2\theta_{13}s^2\theta_{12}\right)+\frac{1}{2}s2\theta_{23}c2\theta_{23}s2\theta_{12}s2\theta_{13}c\theta_{13}\cos\delta\right], \label{reUUUU} \\
\Im\left[U_{\mu 1}U_{\mu 3}^*U_{\tau 1}^*U_{\tau 3}\right]&=&\frac{1}{8}s2\theta_{23}s2\theta_{12}s2\theta_{13}c\theta_{13}\sin\delta.
\label{imUUUU}
\end{eqnarray}
Eq.~(\ref{imUUUU}) is, of course, proportional to the Jarlskog invariant. As discussed above, both Eq.~((\ref{reUUUU}) and Eq.~(\ref{imUUUU}) are invariant under $\theta_{23}\to\theta_{23}+\pi$ or $\theta_{12}\to\theta_{12}+\pi$. They are also invariant under $\theta_{13}\to -\theta_{13}-\pi$. Finally, $\theta_{ij}\to-\theta_{ij}$ accompanied by $\delta\to\delta+\pi$ also leaves both equations unchanged, for all $ij=12,13,23$. 

As in the two flavor case, in order to ``see'' Majorana phases, phenomena involving lepton number violation are required. If all neutrino masses are very small ($\ll 100$~MeV), the rate for neutrinoless double-beta decay is proportional to the magnitude of the $ee$-element of the neutrino mass matrix in the basis where the charged lepton mass matrix and the charged current weak interactions matrix are diagonal:
\begin{eqnarray}
m_{ee}=\sum_{i}U_{ei}^2m_i&=&c_{13}^2\left(c^2_{12}m_1+s^2_{12}m_2e^{2i\phi_2}\right)+s^2_{13}m_3e^{2i(\phi_3+\delta)}, \\
\left|m_{ee}\right|^2&=&\sum_i \left|U_{ei}^2\right|^2 m_i^2+2\sum_{ij=12}^{13,23}\Re\left[U^2_{ei}U^{2*}_{ej}\right]m_im_j.
\end{eqnarray}
It is easy to see that all terms in $|m_{ee}|^2$ depend on products of $s^2\theta_{ij}$ and $c^2\theta_{ij}$ and are hence invariant under $\theta_{ij}\to\theta_{ij}+\pi$ {\sl or} $\theta_{ij}\to-\theta_{ij}$. This implies that all terms are also invariant under $\delta\to\delta+\pi$, which is indeed the case. According to our paraterization, all dependency on the CP-odd phases appears in the form of $\cos(2\delta+2\phi_3)$ and $\cos(2\phi_2-2\delta-2\phi_3)$. These also reveal that $\phi_{2}\to\phi_2+\pi$ or $\phi_3\to\phi_3+\pi$ also leave $|m_{ee}|^2$ invariant, as advertised in Table~\ref{tab:3nus}.

\subsection{The Four-Plus Neutrino Case} \label{subsec:NNu}

In the less familiar case of $N\ge 1$, the lepton mixing matrix is a $3+N\times3+N$ unitary matrix. A general unitary matrix of this size contains $(3+N)^2$ real parameters and at least $3+N$ of those can  be ``field-redefined'' as in the three generation case.  Of the remaining parameters, $(3+N)(2+N)/2$ can be parameterized as mixing angles, while the remaining $(3+N)(2+N)/2$ are CP-odd phases. Finally, of all CP-odd phases, $2+N$ will be defined as Majorana phases $\phi_2,\ldots,\phi_{N+3}$, while the remaining $(2+N)(1+N)/2$ will be denominated Dirac phases.  This would be the end of the story if all sterile neutrino states were coupled to ``sterile charged-fermions'' via the charged-current weak interactions. This not being the case, one is free to rotate the sterile--sterile part of the lepton mixing matrix and further remove $N^2-N$ parameters, $(N-1)(N-2)/2$ CP-odd phases and $N(N-1)/2$ mixing angles.

We choose to parameterize the mixing matrix as a product of all distinct $\mathbf{R^{ij}}(\theta_{ij})$, $i\le 3$, interspersed with the appropriate diagonal phase matrices. Following \cite{Fritzsch:1986gv}, we define Dirac phases so that rotations between the neighboring $i-i+1$ planes are taken to be real, while the rest may ``contain'' a CP violating Dirac phase.  These ``complex rotations'' in the $i-j$ plane refer to the matrix
\begin{equation}
{\bf \widetilde{R}^{ij}}(\theta_{ij},\delta_{ij}) \equiv {\bf
P^i}(\delta_{ij}/2) {\bf P^j}(-\delta_{ij}/2) {\bf
R^{ij}}(\theta_{ij}){\bf P^i}(-\delta_{ij}/2) {\bf
P^j}(\delta_{ij}/2).
\end{equation} 
Finally, we find that we can choose all rotations in the $3-3+N$ plane real. Using these conventions, $U$ will be written as
\begin{equation}
U_{3+N} = 
 \left(\prod_{i=4}^{3+N}\mathbf{R^{3i}}(\theta_{3i}){\bf
\widetilde{R}^{2i}}(\theta_{2i},\delta_{2i}){\bf \widetilde{R}^{1i}}(\theta_{1i},\delta_{1i})\right)
\mathbf{R^{23}}(\theta_{23})\mathbf{\widetilde{R}^{13}}(\theta_{13},\delta)\mathbf{R^{12}}(\theta_{12})
\left(\prod_{i=2}^{3+N}{\bf P^i}(\phi_i)\right), \label{eq:NNuMix}
\end{equation}
where the products are defined as $\prod_{i=1}^n A_i = A_1 A_2 A_3 ... A_n$ and the expression is defined for $N\ge 1$. Several other choices are available \cite{Dooling:1999sg}. Eq.~(\ref{eq:NNuMix}) singles out the ``active'' mixing from the ``sterile'' mixing and is chosen such that the $N=1$ mixing matrix is ``simple,'' as will be described below. 

In the case  $N=1$, Eq.~(\ref{eq:NNuMix}) takes the form
\begin{eqnarray}
&U_4=
\mathbf{R^{34}}(\theta_{34})
\mathbf{\widetilde{R}^{24}}(\theta_{24},\delta_{2})
\mathbf{\widetilde{R}^{14}}(\theta_{14},\delta_{1})
\mathbf{R^{23}}(\theta_{23})\mathbf{\widetilde{R}^{13}}(\theta_{13},\delta)\mathbf{R^{12}}(\theta_{12})
{\bf P^2}(\phi_2){\bf P^3}(\phi_3){\bf P^4}(\phi_4), \\
\nonumber \\
&=\left(\begin{array}{cccc} 
c\theta_{12}c\theta_{13}c\theta_{14} & s\theta_{12}c\theta_{13}c\theta_{14} & s\theta_{13}c\theta_{14}e^{i\delta} & s\theta_{14}e^{i\delta_{1}} \\
\star & \star & s\theta_{23}c\theta_{13}c\theta_{24}-s\theta_{13}s\theta_{14}s\theta_{24}e^{-i(\delta+\delta_{1})} & c\theta_{14}s\theta_{24}e^{i\delta_{2}} \\
\star & \star & c\theta_{23}c\theta_{13}c\theta_{34}-s\theta_{23}c\theta_{13}s{\theta_{24}}s\theta_{34}e^{-i\delta_{2}}-c\theta_{13}s\theta_{14}c\theta_{24}s\theta_{34}e^{-i\delta_{1}} & c\theta_{14}c\theta_{24}s\theta_{34} \\
\star & \star & -c\theta_{23}c\theta_{13}s\theta_{34}-s\theta_{23}c\theta_{13}s\theta_{24}c\theta_{34}e^{-i\delta_{2}}-s\theta_{13}s\theta_{14}c\theta_{24}s\theta_{34}e^{-i(\delta+\delta_{1})} & c\theta_{14}c\theta_{24}c\theta_{34}  \\
\end{array}\right) \nonumber \\
& \times \left(\begin{array}{cccc}
1&0&0&0\\
0&e^{i\phi^2}&0&0\\
0&0&e^{i\phi_3}&0\\
0&0&0&e^{i\phi_4}
\end{array}\right), \label{eq:u4}
\end{eqnarray}
where $\star$s stand for very complicated functions of mixing angles and CP-odd factors that are not illuminating enough to warrant display. One will appreciate that our choice of parameterization is such that $U_{\alpha4}$ elements, for all $\alpha=e,\mu,\tau,s$,  and $U_{ei}$ elements, for all $i=1,2,3,4$, are simple, in parallel with the parameterization of the $3\times 3$ neutrino mixing matrix, Eq.~(\ref{eq:ThreeNuMix}). The mixing matrix is parameterized by twelve real parameters: the six mixing angles $\theta_{12},\theta_{13},\theta_{23},\theta_{14},\theta_{24},\theta_{34}$, the three Dirac phases $\delta,\delta_1,\delta_2$, and the three Majorana phases $\phi_2,\phi_3,\phi_4$.

In the case  $N=2$, Eq.~(\ref{eq:NNuMix}) takes the form
\begin{eqnarray}
U_5&=&
\mathbf{R^{34}}(\theta_{34})
\mathbf{\widetilde{R}^{24}}(\theta_{24},\delta_{24})
\mathbf{\widetilde{R}^{14}}(\theta_{14},\delta_{14})
\mathbf{R^{35}}(\theta_{35})
\mathbf{\widetilde{R}^{25}}(\theta_{25},\delta_{25})
\mathbf{\widetilde{R}^{15}}(\theta_{15},\delta_{15}) \nonumber \\ & \times & 
\mathbf{R^{23}}(\theta_{23})\mathbf{\widetilde{R}^{13}}(\theta_{13},\delta)\mathbf{R^{12}}(\theta_{12})
{\bf P^2}(\phi_2){\bf P^3}(\phi_3){\bf P^4}(\phi_4){\bf P^5}(\phi_5).
\end{eqnarray}
Here the mixing matrix is parameterized in terms of 18 parameters ($5^2-5-(2^2-2)=18$): the nine mixing angles $\theta_{12},\theta_{13},\theta_{23},\theta_{14},\theta_{24},\theta_{34},\theta_{15},\theta_{25},\theta_{35}$, the five Dirac phases $\delta,\delta_{14},\delta_{24},\delta_{15},\delta_{25}$, and the four Majorana phases $\phi_2,\phi_3,\phi_4,\phi_5$. We present approximate expressions for $U_4$ and $U_5$ in Appendix~\ref{app:approx}. One should keep in mind that $U_5$ (and all $U_{3+N}$ matrices with $N\ge 2$) is only defined up to an overall rotation of the sterile sector of the matrix. In the case of $U_5$, one can always rotate around some of the mixing angles and Dirac phases via $U_5\to\mathbf{\widetilde{R}^{45}}(\theta_{45},\delta_{45})U_5$.

The next step is to define the neutrino mass eigenstates. The number of options in the $N=1$ case is large, and we will consider only a few. The most obvious but often more cumbersome choice is to simply order the neutrino masses in ascending label-order: $m_1<m_2<m_3<m_4$. A second choice --- the one we will concentrate on here --- is driven by experimental information currently available regarding active and sterile neutrino masses and mixing angles. Qualitatively, the data tell us that there is one state that is predominantly sterile, while the other three states are predominantly active. The three predominantly active states will be called $\nu_1$, $\nu_2$, and $\nu_3$. These will be ordered as in case C3 of the previous subsection. The predominantly sterile state --- by that we mean the state of maximum $|U_{si}|^2$ --- will be called $\nu_4$. In more detail, $|U_{s1}|^2,|U_{s2}|^2,|U_{s3}|^2$ are constrained in such a way that $|U_{s4}|^2\gtrsim 1/2$. Translating this constraint in terms of mixing angles is not very illuminating --- see case B3 in the three neutrino discussion ---  and will not be discussed here. Generalized versions of this can be applied to the $N>1$ case, where it it is also necessary to specify how the ``mostly sterile'' states are distinguished. A simple choice is to order than in ascending label order, $\it i.e.$, $m_4<m_5<\ldots$.

We now search for symmetries of $U_{3+N}$.  The procedure is entirely analogous to the one presented in the previous subsection, and so are the results. For all ``complex rotations'' ${\bf \widetilde{R}^{ij}}(\theta_{ij},\delta_{ij})$, $\theta_{ij}\to-\theta_{ij}$ and $\delta_{ij}\to\delta_{ij}+\pi$ leave $U_{3+N}$ invariant, and we look for all other redundancies by looking for solutions to 
\begin{equation}
U_{3+N}(\theta'_{ij},\delta'_{ij},\phi'_i)=\mathbf{N_f}U_{3+N}(\theta'_{ij},\delta'_{ij},\phi'_i)\mathbf{N_m},
\end{equation}
where $\mathbf{N_f}$, $\mathbf{N_m}$ are $3+N\times3+N$ diagonal matrices whose elements are all possible combinations of $\pm 1$, with the exception of $[\mathbf{N_m}]_{1,1}\equiv 1$. Such ``sign'' redefinitions of the neutrino mass states and the charged leptons allow all $\theta_{ij}\to\pm(\theta_{ij}+\pi)$ (the $\pm$ sign depends on the specific value of $ij$) regardless of the value of all other parameters. Furthermore, it is easy to show that $\nu_i\to-\nu_i$ identifies all $\phi_i$ with $\phi_i+\pi$. Finally, as in the three neutrino case, $\theta_{ij}\to-\theta_{ij}$ accompanied by $\delta_{ik}\to\delta_{ik}+\pi$ (one $ik$ to each $ij$) also describes the same physics. 

In summary, we are left with the following physical parameter space for $U_{3+N}$, for all values of $N\ge 1$:
\begin{itemize}
\item All Majorana phases $\phi_i\in[0,\pi]$, $i=2,\ldots,3+N$;
\item All Dirac phases $\delta,\delta_{ij}\in[-\pi,\pi[$;
\item All mixing angles can be constrained to the first quadrant: $\cos\theta_{ij},\sin\theta_{ij}>0$.
\end{itemize}
As in the three neutrino case, we choose the Dirac phases to vary along the entire unit circle in the complex plane, while mixing angles only occupy at most the first quadrant. It is important to emphasize that this is a choice. One is also free to, for example, constrain $\delta>0$, while allowing some of the mixing angles to also take negative values \cite{GonzalezGarcia:2001uy}.  

\setcounter{equation}{0} 
\section{Nonstandard Interactions} \label{sec:NonStandardInt}

Some of the results spelled out in the previous section depend on the fact that all neutrino interactions are mediated by Eq.~(\ref{eq:LSM}). New ``weaker-than-weak'' neutrino interactions modify the picture painted in Sec.~\ref{sec:PhyRange} in an interesting way.
 
 Generic non-standard neutrino interactions can be parameterized in a variety of ways. Data constrain new neutrino interactions to be of order or weaker than the weak interactions for neutrino processes that involve momentum transfers less than around one hundred GeV. Assuming that the new neutrino interactions are governed by new particles that weigh more than a few hundred GeV, their physics --- assuming all relevant energies are low enough --- is captured by the following effective Lagrangian, after electroweak symmetry breaking:
 \begin{equation}
 {\cal L}_{\rm NSI} = \frac{1}{\Lambda^2}\left[\left(\bar{\nu}_{\alpha}\xi^{\alpha\beta}\gamma^{\mu}\nu_{\beta}\right){\cal J}_{\mu} + \left(\overline{\nu^c}_{\alpha}\eta^{\alpha\beta}\nu_{\beta}\right){\cal I}\right]+H.c., \label{eq:NSI}
 \end{equation} 
 where 
 ${\cal J}_{\mu}=A_{\ell}^{\alpha\beta}(\bar{\ell}_{\alpha}\gamma_{\mu}\ell_{\beta})+A_{e}^{\alpha\beta}(\bar{e}_{\alpha}\gamma_{\mu}e_{\beta})+A_{uV}^{\alpha\beta}\bar{u}_{\alpha}\gamma_{\mu}u_{\beta}+A_{uA}^{\alpha\beta}\bar{u}_{\alpha}\gamma_{\mu}\gamma_5u_{\beta}+\ldots$ is a generic (axial)vector current made up of quarks and leptons, while ${\cal I}=A_{\ell}^{\alpha\beta}(\bar{e}_{\alpha}\ell_{\beta})+A_{uS}^{\alpha\beta}\bar{u}_{\alpha}u_{\beta}+A_{uP}^{\alpha\beta}\bar{u}_{\alpha}\gamma_5u_{\beta}+\ldots$ is a generic (pseudo)scalar current. $\Lambda$ characterizes the overall strength of the interactions (and does not concern this discussion) while $\xi^{\alpha\beta}$ and $\eta^{\alpha\beta}$ parameterize the neutrino currents. Eq.~(\ref{eq:NSI}) is written in the flavor basis, such that $\alpha,\beta=e,\mu,\tau,s_1,\ldots,s_N$. We have also assumed that neutrinos are Majorana fermions. The $\eta^{\alpha\beta}$ terms violate lepton number and are further constrained by our current understanding of neutrino masses (see, e.g., \cite{de Gouvea:2007xp}). Constraints on $\xi^{\alpha\beta}$ for ``flavor-diagonal'' ${\cal J}_{\mu}$ can be found in \cite{NSI}.
 
As in the previous section, we are interested in the number and range of parameters necessary to describe neutrino related processes. We proceed with the counting in the ``standard'' way. We render all charged-lepton and neutrino masses real and positive. We also perform as many field redefinitions are possible in order to write the charged-current weak interactions as in Eq.~(\ref{eq:LSM}) where the weak and flavor states are related via Eq.~(\ref{def_U}) and $U$ is parameterized as in Eq.~(\ref{U_par}). All other diagonal field redefinitions are employed in order to set $U$ in its standard form. 

We now proceed to ask how $\eta$ and $\xi$ change when one chooses to re-express the neutrino fields in the mass basis. If we assume that all $\eta$ and $\xi$ are known in one specific basis, their values in the other basis should be properly defined. Furthermore, all physically distinguishable couplings should be accessible by the parameterization.

Expressing Eq.~(\ref{eq:NSI}) in the mass basis, we find that $\xi$ and $\eta$ transform as 
\begin{eqnarray}
\xi &\rightarrow& U^\dagger \xi U \\
\nonumber \eta &\rightarrow& U^T \eta U.
\end{eqnarray}
In light of these redefinitions, we revisit Eq.~(\ref{master}), which allowed us to define equivalent classes for the mixing parameters $\theta,\delta,\phi$.  In short hand notation, $U={\bf N_f}U'{\bf N_m}$ and, of course, $U^\dagger \xi U={\bf N_m}(U')^{\dagger}{\bf N_f}\xi{\bf N_f}U'{\bf N_m}$. The same is true for $\eta$ where $U^{\dagger}$ is replaced by $U^T$ since both ${\bf N_{m,f}}$ are real, diagonal matrices (Hermitian). This implies that any constraints lost due to $\xi$ (this will become clear momentarily) will be lost due to $\eta$, regardless of the fact that $\eta$ processes violate lepton number, so it is enough to discuss the transformation of $\xi$.  

The next step in the constraining-the-parameter-space argument is to state that $\theta',\delta',\phi'$ and $\theta,\delta,\phi$ are equivalent because the ${\bf N_{m,f}}$ matrices can be absorbed by sign redefinitions of neutrino and charged-lepton mass eigenstates. The new interactions impose a new constraint: $\theta',\delta',\phi'$ and $\theta,\delta,\phi$ are equivalent only if $\xi$ and ${\bf N_f}\xi{\bf N_f}$ are equivalent, {\it i.e.}, if ${\bf N_f}$ and $\xi$ commute. Hence, if all the elements of $\xi$ are known in one basis, the number of sign redefinitions that can be employed in order to constrain the lepton mixing parameter space is generically reduced.

Revisiting the discussions in Sec.~\ref{sec:PhyRange}, many of the parameter limitations still hold in the face of general new physics, while some no longer apply. Furthermore, in the $N\ge 2$ cases, one needs to revisit the sterile--sterile field redefinition that allowed one to reduce the parameter space by $N^2-N$ parameters. We will not pursue this issue here, and will restrict ourselves to $\xi$ and $\eta$ interactions that involve only active neutrinos.

Restricting ourselves to transformations with $\mathbf{N_{f}=\pm\mathbf{I}}$ within the $3+N$ neutrino scenario, the invariances involving $\theta_{i-1,i}$ and  $-\theta_{i-1,i}$ are no longer available, while the following redundancies remain:
\begin{enumerate}
\item $\phi_i \rightarrow \phi_i +\pi, \forall i>1$;
\item $\theta_{ij} \rightarrow -\theta_{ij}, \delta_{ij} \rightarrow
\delta_{ij} + \pi$ for all ``complex'' $i-j$ rotations;
\item $\theta_{ij} \rightarrow \theta_{ij} + \pi, \theta_{i,i+1} \rightarrow -\theta_{i,i+1}, \theta_{ia} \rightarrow -\theta_{ia}, \theta_{bj} \rightarrow -\theta_{bj}$, for all integers $b<i$ and $i+1 < a < j$.
\end{enumerate}
Item 3 is the $3+N$ neutrino equivalent of the $\mathbf{I},\mathbf{R^{23}}(\pi)$, $\mathbf{-I},\mathbf{P^{2,3}}(\pi)$ lines in Table~\ref{tab:3nus}. Making use of the standard choices and mass-eigenstate redefinitions, we find that while we can restrict all Majorana phases to the $[0,\pi] $ interval and can ignore mixing angle values outside of $[-\pi/2,\pi/2]$,  we can't reduce the parameter space for several of the mixing angles any further without making choices for the elements of $\xi$ and $\eta$.\footnote{Diagonal $\xi$, but not necessarily flavor universal, implies $\xi={\bf N_f}\xi{\bf N_f}$, in which case the ``normal'' parameter space is sufficient.} Ultimately, we find that we can choose all angles associated to complex rotations to be constrained to lie within $[0,\pi/2]$, while the remaining angles are constrained to lie within $[-\pi/2,\pi/2]$. We emphasize that this is a choice. As long as some angles are constrained to lie within $[-\pi/2,\pi/2]$, others can lie in the upper-left quadrant of the unit circle. 
 
 An example is in order. Restricting ourselves to the two flavor case, the discussion above implies that, in the presence of non-standard flavor changing interactions, all physically distinguishable scenarios can be parameterized if $\phi\in[0,\pi]$ and $\theta\in[-\pi/2,\pi/2]$, such that the {\sl sign} of $\theta$ is physically distinguishable. Further consider non-standard interactions where ${\cal J}_{\mu}=\bar{e}\gamma_{\mu}e$, ${\cal I}= 0$. In this case, non-standard neutrino interactions affect the coherent neutrino--electron scattering scattering amplitudes that govern the matter effects in neutrino oscillations. In order to compute neutrino oscillation probabilities, it is convenient to express these amplitudes in the neutrino mass basis:
\begin{equation}
\mathcal{A}_{ij} = F_Z\delta_{ij} + F_W U_{ei}^* U_{ej} +
F_{\rm NSI}U^*_{\alpha i}\xi^{\alpha \beta} U_{\beta j} ,
\end{equation}
where $F$s are functions containing process-dependent kinematical factors and coupling constants.  For our purposes, it is enough to know that these functions are independent of neutrino mixing parameters and couplings. The first two terms correspond to $Z$ and $W$ exchange standard model diagrams, respectively, while the third term contains the effects of the new interactions. In matrix form, 
\begin{eqnarray}
&{\cal A}_{\rm mass}=\left(F_Z+\frac{F_W}{2}+\frac{F_{\rm NSI}}{2}\left(\xi_{ee}+\xi_{\mu\mu}\right)\right) \left(\begin{array}{cc} 1 & 0 \\ 0 & 1\end{array}\right) + 
\left(\frac{F_W}{2}+\frac{F_{\rm NSI}}{2}\left(\xi_{ee}-\xi_{\mu\mu}\right)\right) \left(\begin{array}{cc} c2\theta & s2\theta e^{i\phi} \\ s2\theta e^{-i\phi} & -c2\theta \end{array}\right) + \nonumber \\ 
&\frac{F_{\rm NSI}}{2}
\left(\begin{array}{cc} -\left(\xi_{\mu e}+\xi_{e\mu}\right)s2\theta & \left[\left(\xi_{e\mu}-\xi_{\mu e}\right)+\left(\xi_{e\mu}+\xi_{\mu e}\right)c2\theta \right]e^{i\phi}\\ \left[\left(\xi_{\mu e}-\xi_{e\mu}\right)+\left(\xi_{e\mu}+\xi_{\mu e}\right)c2\theta\right]e^{-i\phi} & \left(\xi_{\mu e}+\xi_{e\mu}\right)s2\theta \end{array}\right).
\end{eqnarray}
It is important to appreciate that only the magnitude of the different elements of ${\cal A}_{\rm mass}$ are relevant. This is best seen by analyzing the $F_W$ term, which is present in the Standard Model and drives ``normal'' matter effects in neutrino oscillations. On the other hand, the relative sign of different terms within an element does matter. With this is mind, it is easy to see that $\theta$ and $\theta+\pi$ describe the same physics. This is not the case of $\theta$ and $-\theta$. This is visible in all flavor off-diagonal NSI terms. Also, as expected, the Majorana phase $\phi$ plays no role in this observable. Finally, note that if $\xi$ were diagonal in the flavor basis, $\theta$ and $-\theta$ would describe the same phenomenon, and the only impact of this NSI would be to modify the strength of the $W$-mediated matter effect.

In the case of three active neutrinos, one of the three mixing angles can be restricted to $[0,\pi/2]$ while the other two have an expanded parameters space. One concrete choice is as follows. For generic non-standard interactions, 
\begin{itemize}
\item $\theta_{12},\theta_{23}\in[-\pi/2,\pi/2]$, $\theta_{13}\in[0,\pi/2]$;
\item $\delta\in [-\pi,\pi[$;
\item $\phi_2,\phi_3\in[0,\pi]$.
\end{itemize}

Before concluding, we would like to emphasize that the discussion above assumes that all elements of $\xi$, including their signs, are known in some basis. If this is not the case, we can pick all $\theta_{ij}$ positive and allow both signs for the off-diagonal elements of $\xi$. This fact was already emphasized in \cite{Fornengo:2001pm}. In the two-neutrino case discussed above, if $\xi$ were measured via anomalous matter effects, one would not be able to tell whether $\xi_{\mu e}$ has a particular sign and $\theta$ is negative, or whether $\xi_{\mu e}$ has the opposite sign and $\theta$ is negative. On the other hand, the sign of all elements of $\xi$ should be, at least in principle, observable, as long as one can measure observables that depend on the interference between standard model diagrams and those involving $\xi$.

\section{Conclusion} \label{sec:Conclusion}

It is well-known that while quark mixing phenomena are governed by a three-by-three unitary matrix, all physically distinguishable mixing scenarios are parameterized by three mixing angles constrained to lie within $[0,\pi/2]$ and one CP-odd phase that lies within the entire unit circle $[-\pi,\pi[$ \cite{Yao:2006px}. 

Assuming that only three active neutrinos exist, the situation in the lepton sector is similar but not identical.  All physically distinguishable lepton mixing scenarios are parameterized by three mixing angles constrained to lie within the first quadrant, one Dirac CP-odd phase that lies within the entire unit circle $\delta\in[-\pi,\pi[$, and two Majorana CP-odd phases that can be chosen to lie in the upper half of the unit circle, $\phi_i\in[0,\pi]$. We argued that the full parameter range for some of the mixing angles can only be properly defined once the neutrino mass eigenstates $\nu_1$, $\nu_2$, and $\nu_3$ are unambiguously defined. In the standard definition of neutrino mass eigenstates, all three mixing angles are constrained to lie within $\theta_{ij}\in [0,\pi/2]$. In other equally unambiguous but more cumbersome definitions (see case B3) smaller intervals suffice.  

In the presence of sterile neutrinos, the parameter space grows in a nontrivial way (from 6 parameters in the three neutrino world to $(3+N)^2-(3+N)-(N^2-N)$ in a world with N sterile neutrinos). We find, however, that the physical range for the mixing parameters can be chosen in a way that mimics the standard three neutrino parameter space: all mixing angles can be chosen at most between $[0,\pi/2]$ as long as all Dirac CP-odd phases are allowed to vary within $[-\pi,\pi[$. On the other hand, all Majorana phases $\phi_i$, $i=2,\ldots,3+N$ can always be constrained to lie within $[0,\pi]$. 

If neutrinos participate in new ``weaker-than-weak'' interactions, the treatment of the  neutrino mixing parameters needs to be reexamined. In particular we find that in the presence of new interactions which are off-diagonal in the flavor basis mixing angles within $]0,\pi/2]$ and those within $[-\pi/2,0[$ can be interpreted as describing different phenomena, and the physical neutrino parameter space is doubled. It is instructive to describe a very simple  two-flavor example to illustrate what this means. Define the mass eigenstates such that $\Delta m^2>0$ and assume that $|\theta|\to(\pi/4)_-$ and that the Majorana phase vanishes.\footnote{In the case of maximum mixing, one needs to worry about how to define the two different mass eigenstates. One way is to use the sign that is being discussed here, in which case the answer to the question we are asking is a matter of definition! For the discussion here, one can assume that $|\theta|$ is very close to $\pi/4$ but not identical, such that the definition of the two mass eigenstates is unambiguous. All results presented below are obtained under this assumption.} In this case,
\begin{eqnarray}
\nu_e=\frac{1}{\sqrt{2}}\left(\nu_1\pm\nu_2\right), \label{eq:nue_pi/4}\\
\nu_{\mu}=\frac{1}{\sqrt{2}}\left(\mp \nu_1+\nu_2\right),
\end{eqnarray}
where the ambiguous sign depends on whether $\theta=\pi/4$ or $-\pi/4$. A reasonable question to ask is whether the electron is the ``symmetric'' or the ``antisymmetric'' linear combination of $\nu_1$ and $\nu_2$. In the absence of new interactions, the answer is that we can't tell. Both sign choices are physically equivalent. Now add to the Lagrangian a new interaction of the type discussed in Sec.~\ref{sec:NonStandardInt}, where $\xi_{\mu e}=\xi_{e\mu}$ and all other $\xi$ values vanish in the flavor basis. In this case, the oscillation frequency through an electron background is modified from $\Delta\equiv \Delta m^2/(2E)$ ( $P_{e\mu}\propto\sin^2(\Delta L/2)$) to
\begin{equation}
\Delta_{\rm matter}^2=\Delta^2+F_W^2+4F_{\rm NSI}^2\xi_{\mu e}^2\pm4\Delta(F_{\rm NSI}\xi_{\mu e}).
\end{equation}
Here, the $\pm$ sign is the same as the $\pm$ sign that appears in the electron-neutrino definition, Eq.~(\ref{eq:nue_pi/4}). In this case, a measurement of $\Delta_{\rm matter}$ should reveal whether $\nu_e\propto \nu_1+\nu_2$ or $\nu_e\propto\nu_1-\nu_2$. This statement is only true if the sign of $F_{\rm NSI}\xi_{\mu e}$ is unambiguously determined, and the measurement of $\Delta_{\rm matter}$ does not do it independently. In principle, it should be possible to measure the sign of $\xi_{\mu e}$, although it may be very challenging in practice.

Our purpose in performing this exercise was to systematize the procedure behind the definition of the range for neutrino mixing parameters. Along the way, we expanded well-known (but sometimes misinterpreted or forgotten) results concerning the mixing angles and Dirac phase in the three active neutrino case to more than three neutrinos, and made explicit the origin for the $[0,\pi]$ range for the Majorana phases. We also emphasized the importance of properly defining the neutrino mass eigenstates before defining the physical range of the neutrino mixing parameters.

\setcounter{equation}{0} \setcounter{footnote}{0}
\appendix
\section{Notation}
\label{app:notation}

 Neutrino mixing and rephasing invariances can be
expressed as products of two distinct types of matrices, plus the identity ${\bf I}$.\footnote{One may use
$U(n)$ algebras to parameterize the $n\times n$ mixing matrix. 
These algebras are well-known, but their use would require different
commutation relations for different values of $n$, which is not conducive
to a ``$n$-independent'' description.} These are real orthogonal rotations in the $a-b$ plane,
${\bf R^{ab}}(\theta)$, given by 
\begin{equation}
{\bf R^{ab}}(\theta) \equiv \left\{ \begin{array}{l}
\left[\mathbf{R^{ab}}(\theta)\right]_{aa}=\left[\mathbf{R^{ab}}(\theta)\right]_{bb}=\cos\theta, \\
\left[[\mathbf{R^{ab}}(\theta)\right]_{ab}=-\left[\mathbf{R^{ab}}(\theta)\right]_{ba}=\sin\theta, \\
\left[\mathbf{R^{ab}}(\theta)\right]_{ij}=\delta_{ij},~~~~~{ij\neq ab}
\end{array}
\right. \label{eq:Rab}
\end{equation}
and single diagonal phase rotations ${\bf P^{a}}(\phi)$, given by
\begin{equation}
[{\bf P^{a}}(\phi)]_{ij} = \delta_{ij} e^{i\phi \delta_{ia}}.
\label{eq:Pa}
\end{equation}
Furthermore, discrete permutations of $a\leftrightarrow b$ 
elements of vectors are accomplished with ${\bf S^{ab}} = {\bf R^{ab}}(\pi/2){\bf P^a}(\pi)$.
It is a simple matter to write an arbitrary mixing matrix with such components, as done extensively in the text (Eqs.~(\ref{eq:TwoNuMix},\ref{eq:ThreeNuMix},\ref{eq:NNuMix})).

 The matrices above satisfy the following commutation relations:
\begin{eqnarray}
\left[{\bf R^{ab}}(\theta),{\bf R^{bc}}(\theta^\prime) \right] &=&
\sin\theta\left(1-\cos\theta^\prime\right){\bf A^{ab}} +
\sin\theta^\prime\left(1-\cos\theta\right){\bf A^{bc}} +
\frac{1}{2}\sin\theta \sin\theta^\prime \left\{ {\bf
R^{ac}}\left(-\frac{\pi}{2}\right) - {\bf
R^{ac}}\left(\frac{\pi}{2}\right) \right\},\\
\left[ {\bf R^{ab}}(\theta),{\bf P^a}(\phi) \right] &=& \sin\theta
\left(e^{i\phi}-1\right) {\bf A^{ab}},\\
\left[ {\bf R^{ab}}(\theta),{\bf A^{ab}} \right] &=& -\sin\theta
\left\{ {\bf P^b}(\pi) - {\bf P^a}(\pi) \right\},\\
\left[ {\bf R^{ab}}(\theta),{\bf A^{ac}} \right] &=& \sin\theta {\bf
A^{bc}} + \frac{1}{2}\left( \cos\theta - 1\right)\left\{ {\bf
R^{ac}}\left(-\frac{\pi}{2}\right) - {\bf R^{ac}}\left(\frac{\pi}{2}\right) \right\},\\
\left[ {\bf A^{ab}},{\bf P^{a}}(\phi) \right] &=&
\frac{1}{2}\left(e^{i\phi} - 1\right)\left\{ {\bf
R^{ab}}\left(-\frac{\pi}{2}\right) - {\bf R^{ab}}\left(\frac{\pi}{2}\right) \right\},\\
\left[ {\bf A^{ab}},{\bf A^{ac}} \right] &=& \frac{1}{2}\left\{ {\bf
R^{bc}}\left(-\frac{\pi}{2}\right) - {\bf R^{bc}}\left(\frac{\pi}{2}\right) \right\},
\end{eqnarray}
while all other commutators vanish.  This matrix set closes upon itself with the addition of 
the discrete transformation
 \begin{equation}
 {\bf A^{ab}} = {\bf A^{ba}} = \frac{1}{2} {\bf
 R^{ab}}\left(\frac{\pi}{2}\right)
 \left\{ {\bf P^b}(\pi) - {\bf P^a}(\pi) \right\}.
\end{equation}
The following identities are easy to prove and were used extensively to derive the results presented in the body of this paper:
\begin{eqnarray}
{\bf P^a}(\phi){\bf P^a}(-\phi) &=& \mathbf{I}, \\
{\bf R^{ab}}(\theta){\bf R^{ab}}(-\theta) &=& {\bf R^{ab}}(\theta){\bf R^{ba}}(\theta) = \mathbf{I}, \\
{\bf R^{ab}}(\theta + \theta^\prime) &=& {\bf R^{ab}}(\theta){\bf R^{ab}}(\theta^\prime), \\
{\bf P^{a}}(\phi + \phi^\prime) &=& {\bf P^{a}}(\phi){\bf P^{a}}(\phi^\prime), \\
 {\bf P^a}(\pi){\bf P^b}(\pi) &=& {\bf R^{ab}}(\pi), \\
 {\bf P^b}(\pi) {\bf R^{ab}}(\theta) &=& {\bf R^{ab}}(-\theta) {\bf P^b}(\pi), \\
 {\bf R^{ab}}(\theta){\bf P^a}(\phi) - {\bf P^b}(\phi){\bf
R^{ab}}(\theta) &=& \cos\theta \left(e^{i\phi}-1\right) {\bf
A^{ab}}, \\
{\bf A^{ab}P^a}(\phi) &=& {\bf P^b}(\phi){\bf A^{ab}} , \\
{\bf R^{ab}}(\theta){\bf R^{ca}}\left(\frac{\pi}{2}\right) &=& {\bf
R^{ca}}\left(\frac{\pi}{2}\right){\bf R^{bc}}(\theta).
\end{eqnarray}

\section{Dirac CP conservation and negative $\theta_{13}$}
\label{neg_ue3}
In the three active neutrino case, all neutrino oscillation CP-violating phenomena are governed by the Dirac CP-odd phase $\delta$ --- the impact of Majorana phases is minuscule and can be trivially dismissed. CP conservation in neutrino oscillations would imply that $e^{i\delta}$ is real and allow a significant reduction of the lepton mixing parameter space. According to the standard choice of parameter ranges, real $e^{i\delta}$ corresponds to $\delta=0$ or $\delta=\pi$, such that Dirac CP-conservation implies that the original six-dimensional mixing parameter space splits into two disjoint five-dimensional ones. 

This split parameter space can be avoided by choosing a different range for $\delta$ and $\theta_{13}$. If one restrains $\delta\in[-\pi/2,\pi/2]$, $\theta_{13}\in[-\pi/2,\pi/2]$ is required to specify all physically distinguishable mixing matrices (we are assuming mass-definition scheme case C3). The physical range for all other mixing parameters is unchanged.  This parameterization was advocated in \cite{Latimer:2004hd} using simalar symmetry arguments.  In this case vacuum (or matter effected) neutrino oscillation searches can detect the sign of $\theta_{13}$ \cite{Latimer:2004gz,Fogli:2005cq}, or equivalently the $\delta \in [-\pi/2,\pi/2]$ versus $\delta \in \pm[\pi/2,\pi]$ value.

\begin{figure}[t]
\begin{center}
\includegraphics[scale=.50]{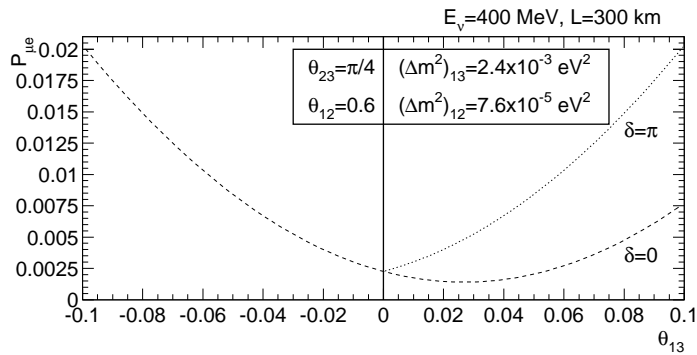}
\caption{$P_{\mu e}$ in vacuum as a function of $\theta_{13}$, assuming no Dirac CP conservation. The dashed (dotted) curve corresponds to $\delta=0$ ($\delta=\pi$). If both values of $\delta$ are part of the physical parameter space, only positive values of $\theta_{13}$ are required to parameterized all distinct phenomena. On the other hand, if one also allows for negative values of $\theta_{13}$, Dirac CP conservation is uniquely determined by $\delta=0$, while $\delta=\pi$ is not part of the physical parameter space.}
\label{fig:cpcons}
\end{center}
\end{figure}

Fig.~\ref{fig:cpcons} illustrates these two choices for the $\theta_{13}$ parameter ranges, assuming Dirac CP conservation. It depicts $P_{\mu e}$ in vacuum as a function of $\theta_{13}$, assuming no Dirac CP conservation, for $L=300$~km and $E_{\nu}=400$~MeV. $\theta_{23}$, $\theta_{12}$ and the neutrino mass-squared differences are fixed at values consistent with the current neutrino oscillation data \cite{NeutrinoReview}.The dashed (dotted) curve corresponds to $\delta=0$ ($\delta=\pi$). If one allows for negative values of $\theta_{13}$, Dirac CP conservation is uniquely determined by $\delta=0$ --- one can trivially see that the values of $P_{\mu e}$ associated to $\delta=\pi$ and $\theta_{13}>0$ are reproduced for $\delta=0$ and $-\theta_{13}<0$.

\section{Approximate Expressions for $U_4$ and $U_5$}
\label{app:approx}

We chose to parameterize the $(3+N)\times(3+N)$ neutrino mixing matrix according to Eq.~(\ref{eq:NNuMix}). Sec.~\ref{subsec:NNu} contains an incomplete expression for $U_4$, Eq.~(\ref{eq:u4}). Here we present approximate expressions for both $U_4$ and $U_5$ which we hope will not only serve those interested in sterile neutrino mixing but also illustrate what our parameterization ``looks like'' in the case of more than one sterile neutrino.

We assume that all mixing angles except $\theta_{23}$ and $\theta_{12}$ are small. Such an approximation is in good agreement with  the current world neutrino data \cite{NeutrinoReview}. The linearized expression for $U_4$ is 
\begin{eqnarray}
& U_4= \left(\begin{array}{cc}U_3 & \Theta_1 \\ \varTheta_1^t & 1\end{array}\right)\left(\begin{array}{cccc}
1&0&0&0\\
0&e^{i\phi^2}&0&0\\
0&0&e^{i\phi_3}&0\\
0&0&0&e^{i\phi_4}
\end{array}\right),~~~{\rm where} \\
& U_3 = \left(\begin{array}{ccc} 
c\theta_{12} & s\theta_{12} & \theta_{13}e^{i\delta} \\
-s\theta_{12}c\theta_{23}-c\theta_{12}s\theta_{23}\theta_{13}e^{-i\delta} & c\theta_{12}c\theta_{23}-s\theta_{12}s\theta_{23}\theta_{13}e^{-i\delta} & s\theta_{23}  \\
s\theta_{12}s\theta_{23}-c\theta_{12}c\theta_{23}\theta_{13}e^{-i\delta} & -c\theta_{12}s\theta_{23}-s\theta_{12}c\theta_{23}\theta_{13}e^{-i\delta} & c\theta_{23} \end{array}\right), \label{eq:u3}\\
& \Theta_1 = \left(\begin{array}{c} 
 \theta_{14}e^{i\delta_{1}} \\
 \theta_{24}e^{i\delta_{2}} \\
 \theta_{34} 
\end{array}\right),~~~\varTheta_1=\left(\begin{array}{c} 
-c\theta_{12}\theta_{14}e^{-i\delta_1}+s\theta_{12}c\theta_{23}\theta_{24}e^{-i\delta_2}-s\theta_{12}s\theta_{23}\theta_{34} \\\ -s\theta_{12}\theta_{14}e^{-i\delta_1}-c\theta_{12}c\theta_{23}\theta_{24}e^{-i\delta_2}+c\theta_{12}s\theta_{23}\theta_{34} \\ -c\theta_{23}\theta_{34}-s\theta_{23}\theta_{24}e^{-i\delta_{2}} 
\end{array}\right).
\end{eqnarray}

The linearized expression for  $U_5$ is
\begin{eqnarray}
& U_5= \left(\begin{array}{cc}U_3 & \Theta_2 \\ \varTheta_2^t & \mathbf{I} \end{array}\right)\left(\begin{array}{ccccc}
1&0&0&0&0\\
0&e^{i\phi^2}&0&0&0\\
0&0&e^{i\phi_3}&0&0\\
0&0&0&e^{i\phi_4}&0\\
0&0&0&0&e^{i\phi_5}
\end{array}\right),~~~{\rm where} \label{eq:u5}\\
& \mathbf{I} = \left(\begin{array}{cc} 
1 & 0 \\
0 & 1 \end{array}\right),~~~~\Theta_2 = \left(\begin{array}{cc} 
 \theta_{14}e^{i\delta_{14}} & \theta_{15}e^{i\delta_{15}} \\
 \theta_{24}e^{i\delta_{24}} & \theta_{25}e^{i\delta_{25}} \\
 \theta_{34} &  \theta_{35} 
\end{array}\right), \\
& \varTheta_2=\left(\begin{array}{cc} 
-c\theta_{12}\theta_{14}e^{-i\delta_{14}}+s\theta_{12}c\theta_{23}\theta_{24}e^{-i\delta_{24}}-s\theta_{12}s\theta_{23}\theta_{34} & -c\theta_{12}\theta_{15}e^{-i\delta_{15}}+s\theta_{12}c\theta_{23}\theta_{25}e^{-i\delta_{25}}-s\theta_{12}s\theta_{23}\theta_{35} \\
-s\theta_{12}\theta_{14}e^{-i\delta_{14}}-c\theta_{12}c\theta_{34}\theta_{24}e^{-i\delta_{24}}+c\theta_{12}s\theta_{23}\theta_{34} & -s\theta_{12}\theta_{15}e^{-i\delta_{15}}-c\theta_{12}c\theta_{34}\theta_{25}e^{-i\delta_{25}}+c\theta_{12}s\theta_{23}\theta_{35} \\
-s\theta_{23}\theta_{24}e^{-i\delta_{24}}-c\theta_{23}\theta_{34} & -s\theta_{23}\theta_{25}e^{-i\delta_{25}}-c\theta_{23}\theta_{35}
\end{array}\right), \nonumber \\
\end{eqnarray}
and $U_3$ is given by Eq.~(\ref{eq:u3}).

It is easy to imagine how Eq.~(\ref{eq:u5}) generalizes to the the case of $N=3$ or more sterile neutrinos. 
 
\section*{Acknowledgments}
This work was inspired by a question raised by Florian Plentinger concerning the physical range for Majorana phases. AdG would like to thank the hospitality of the Physics Department at Columbia University, where this work was completed.
This work is sponsored in part by the US Department of Energy Contract DE-FG02-91ER40684.


\begin{thebibliography}{99}

\bibitem{NeutrinoReview}
See, for example,
M.~C.~Gonzalez-Garcia and M.~Maltoni,
  arXiv:0704.1800 [hep-ph];
A.~Strumia and F.~Vissani,
  arXiv:hep-ph/0606054;
 R.~N.~Mohapatra {\it et al.},
  Rept.\ Prog.\ Phys.\  {\bf 70}, 1757 (2007);
 A.~de Gouv\^ea,
 Mod.\ Phys.\ Lett.\  A {\bf 19}, 2799 (2004);
 A.~de Gouv\^ea,
 arXiv:hep-ph/0411274.

\bibitem{Cabibbo:1963yz}
  N.~Cabibbo,
  Phys.\ Rev.\ Lett.\  {\bf 10}, 531 (1963).

\bibitem{Branco:1999fs}
  G.C.~Branco, L.~Lavoura and J.P.~Silva, {\it CP violation},
Oxford University Press, New York (1999).

\bibitem{SeeSaw} P.~Minkowiski, Phys.\ Lett.\ B {\bf 67}, 421 (1977);
M. Gell-Mann, P. Ramond and R. Slansky in {\it Supergravity}, eds.
D. Freedman and P. Van Niuenhuizen (North Holland, Amsterdam, 1979),
p.~315; T. Yanagida in {\it Proceedings of the Workshop on Unified
Theory and Baryon Number in the Universe}, eds. O.~Sawada and
A.~Sugamoto (KEK, Tsukuba, Japan, 1979); S.L.~Glashow, {\it 1979
Carg\`ese Lectures in Physics --- Quarks and Leptons}, eds. M.~L\'evy
{\it et al.} (Plenum, New York, 1980), p.~707. See also R.N. Mohapatra and G.
Senjanovi\'c, Phys.\ Rev.\ Lett.\ {\bf 44}, 912 (1980) and \cite{Schechter:1980gr}.

\bibitem{low_seesaw}
A.~de Gouv\^ea,
  Phys.\ Rev.\  D {\bf 72}, 033005 (2005).
See T.~Asaka, S.~Blanchet and M.~Shaposhnikov,
  Phys.\ Lett.\  B {\bf 631}, 151 (2005);
T.~Asaka and M.~Shaposhnikov,
  Phys.\ Lett.\  B {\bf 620}, 17 (2005);
  A.~de Gouv\^ea, J.~Jenkins and N.~Vasudevan,
  Phys.\ Rev.\  D {\bf 75}, 013003 (2007);
  M.~Shaposhnikov,
  Nucl.\ Phys.\  B {\bf 763}, 49 (2007);
F.L.~Bezrukov and M.~Shaposhnikov,
  Phys.\ Rev.\  D {\bf 75}, 053005 (2007);
  A.~de Gouv\^ea,
  arXiv:0706.1732 [hep-ph] 
for recent discussions of the phenomenology of low-energy versions of the seesaw mechanism.

\bibitem{maj_CP_phase}
S.M.~Bilenky, J.~Hosek and S.T.~Petcov,
  Phys.\ Lett.\  B {\bf 94}, 495 (1980);
  M.~Doi, T.~Kotani, H.~Nishiura, K.~Okuda and E.~Takasugi,
  Phys.\ Lett.\  B {\bf 102}, 323 (1981). 
See also B.~Kayser,
  Adv.\ Ser.\ Direct.\ High Energy Phys.\  {\bf 3}, 334 (1989).


\bibitem{Schechter:1980gr}
  J.~Schechter and J.W.F.~Valle,
  Phys.\ Rev.\  D {\bf 22}, 2227 (1980).


\bibitem{Fogli:1996ne}
  G.L.~Fogli, E.~Lisi and D.~Montanino,
  Phys.\ Rev.\  D {\bf 54}, 2048 (1996).

\bibitem{de Gouvea:2000cq}
  A.~de Gouv\^ea, A.~Friedland and H.~Murayama,
  Phys.\ Lett.\  B {\bf 490}, 125 (2000).

\bibitem{GonzalezGarcia:2000sq}
  M.C.~Gonzalez-Garcia, M.~Maltoni, C.~Pe\~na-Garay and J.W.F.~Valle,
  Phys.\ Rev.\  D {\bf 63}, 033005 (2001).

\bibitem{DeGouvea:2001ag}
  A.~de Gouv\^ea,
  Nucl.\ Instrum.\ Meth.\  A {\bf 503}, 4 (2001).

\bibitem{Fornengo:2001pm}
  N.~Fornengo, M.~Maltoni, R.~T.~Bayo and J.W.F.~Valle,
  Phys.\ Rev.\  D {\bf 65}, 013010 (2002).

\bibitem{Latimer:2004hd}
  D.~C.~Latimer and D.~J.~Ernst,
  Phys.\ Rev.\ D {\bf 71}, 017301 (2005).

\bibitem{Yao:2006px}
  W.~M.~F.~Yao {\it et al.}  [Particle Data Group],
  J.\ Phys.\ G {\bf 33}, 1 (2006).

\bibitem{Jenkins:2007ip}
  E.E.~Jenkins and A.V.~Manohar,
  Nucl.\ Phys.\  B {\bf 792}, 187 (2008).

\bibitem{Haba:2000be}
  N.~Haba and H.~Murayama,
  Phys.\ Rev.\  D {\bf 63}, 053010 (2001).

\bibitem{deGouvea:2002gf}
  A.~de Gouv\^ea, B.~Kayser and R.N.~Mohapatra,
  Phys.\ Rev.\  D {\bf 67}, 053004 (2003).

\bibitem{nu_nubar}
L.F.~Li and F.~Wilczek,
  Phys.\ Rev.\  D {\bf 25}, 143 (1982).
  J.~Schechter and J.W.F.~Valle,
  Phys.\ Rev.\  D {\bf 23}, 1666 (1981);
  J.~Bernab\'eu and P.~Pascual,
  Nucl.\ Phys.\  B {\bf 228}, 21 (1983);
P.~Langacker and J.~Wang,
  Phys.\ Rev.\  D {\bf 58}, 093004 (1998).


\bibitem{Petcov:2001sy}
  This choice of masses was pursued, for example, in S.T.~Petcov and M.~Piai,
  Phys.\ Lett.\  B {\bf 533}, 94 (2002).

\bibitem{Gluza:2001de}
J.~Gluza and M.~Zralek,
Phys.\ Lett.\ B {\bf 517}, 158 (2001).

\bibitem{Fritzsch:1986gv}
  H.~Fritzsch and J.~Plankl,
  Phys.\ Rev.\  D {\bf 35}, 1732 (1987).

\bibitem{Dooling:1999sg}
See, for example, 
  D.~Dooling, C.~Giunti, K.~Kang and C.W.~Kim,
  Phys.\ Rev.\  D {\bf 61}, 073011 (2000).

\bibitem{GonzalezGarcia:2001uy}
  M.C.~Gonzalez-Garcia, M.~Maltoni and C.~Pe\~na-Garay,
  Phys.\ Rev.\  D {\bf 64}, 093001 (2001).

\bibitem{de Gouvea:2007xp}
  A.~de Gouv\^ea and J.~Jenkins,
  Phys.\ Rev.\  D {\bf 77}, 013008 (2008).


\bibitem{NSI}
For recent discussions see, for example,
J.~Barranco, O.G.~Miranda, C.A.~Moura and J.W.F.~Valle,
  Phys.\ Rev.\  D {\bf 73}, 113001 (2006)
  [arXiv:hep-ph/0512195].
A.~Friedland, C.~Lunardini and M.~Maltoni,
  Phys.\ Rev.\  D {\bf 70}, 111301 (2004);
A.~Friedland, C.~Lunardini and C.~Pena-Garay,
  Phys.\ Lett.\  B {\bf 594}, 347 (2004);
S.~Davidson, C.~Pe\~na-Garay, N.~Rius and A.~Santamaria,
  JHEP {\bf 0303}, 011 (2003);
A.M.~Gago, M.M.~Guzzo, P.C.~de Holanda, H.~Nunokawa, O.L.G.~Peres, V.~Pleitez and R.~Zukanovich Funchal,
  Phys.\ Rev.\  D {\bf 65}, 073012 (2002).

\bibitem{Latimer:2004gz}
  D.C.~Latimer and D.~J.~Ernst,
  Phys.\ Rev.\ C {\bf  71}, 062501 (2005).

\bibitem{Fogli:2005cq}
  G.L.~Fogli, E.~Lisi, A.~Marrone and A.~Palazzo,
  Prog.\ Part.\ Nucl.\ Phys.\ {\bf 57}, 742 (2006).



 \end{thebibliography}
 \end{document}